\documentclass[
	aps,pra,
	floatfix,
	reprint,
	amsmath,amssymb,
	longbibliography
]{revtex4-2}
\usepackage{graphicx}
\usepackage{mathtools}
\usepackage{pst-all,pstricks-add}

\newcommand{\expv}[1]{\langle #1 \rangle}           %Create a expectation value with an argument
\newcommand{\myvector}[1]{\boldsymbol{\mathrm{#1}}}
\newcommand{\unitvec}[1]{\boldsymbol{\hat{\mathrm{#1}}}}

\newcommand{\rhoi}{\hat{\rho}_\mathrm{i}}

\newcommand{\rhoo}{\hat{\rho}_\mathrm{0}}
\newcommand{\rhooorder}[1]{\hat{\rho}^{(#1)}_0}
\newcommand{\rhoiorder}[1]{\hat{\rho}^{(#1)}_\mathrm{i}}
\newcommand{\rhoforder}[1]{\hat{\rho}^{(#1)}_\mathrm{f}}

\newcommand{\rhof}{\hat{\rho}_\mathrm{f}}
\newcommand{\rhomeas}{\hat{\rho}_\mathrm{m}}

\newcommand{\rhofderivorder}[1]{\hat{\dot{\rho}}^{(#1)}_\mathrm{f}}

\newcommand{\rinitial}{\myvector{r}_\textrm{0}}
\newcommand{\rinitiali}{\myvector{r}_\textrm{$i$}}
\newcommand{\rinitialperp}{\myvector{r}_{\textrm{0}\perp}}

\newcommand{\rfinal}{\myvector{r}_\textrm{f}}

\newcommand{\Horder}[1]{H^{(#1)}}

\newcommand{\Hsopt}{H_\textrm{s opt}}
\newcommand{\Hcorr}{H_\textrm{corr}}
\newcommand{\Hclean}{H_\textrm{clean}}

\newcommand{\score}{\hat{L}}

\newcommand{\iop}{\hat{I}}
\newcommand{\sigmax}{\hat{\sigma}_x}
\newcommand{\sigmay}{\hat{\sigma}_y}
\newcommand{\sigmaz}{\hat{\sigma}_z}
\newcommand{\sigmagen}[1]{\hat{\sigma}_{\boldsymbol{\mathrm{#1}}}}
\newcommand{\sigmavec}{\boldsymbol{\hat{\sigma}}}
\newcommand{\lambdaest}{\lambda_\mathrm{est}}

\newcommand{\channel}{\hat{\Gamma}}
\newcommand{\channellambda}{\hat{\Gamma}(\lambda)}

\newcommand{\uprep}{\hat{U}_\mathrm{prep}}
\newcommand{\uc}{\hat{U}_{\boldsymbol{\mathrm{c}}}}

\DeclareMathOperator{\Trace}{Tr}
\DeclareMathOperator{\variance}{var}

\begin{document}

\author{David Collins}
\affiliation{Department of Physical and Environmental Sciences, Colorado Mesa University, Grand Junction, CO 81501}
\email{dacollin@coloradomesa.edu}
%\thanks{Author to whom correspondence should be addressed.}

\author{Taylor Larrechea} 
\affiliation{Vivint, Lehi, UT 84043}
%\email{}

\title{Qubit-channel metrology with highly noisy initial states and additional undesirable noisy evolution}

\begin{abstract}
  We consider protocols for estimating the parameter in a single-parameter unital qubit channel, assuming that the available initial states are highly mixed with very low purity. We compare two protocols, each invoking the channel once, via their quantum Fisher informations. One uses $n$ qubits prepared in a particular correlated input state and subsequently queries the channel on one qubit. The other uses a single qubit. We extend the results of Collins~\cite{collins19} by allowing for additional noisy evolution on the spectator qubits in the $n$-qubit protocol. We provide simple algebraic expressions that will determine when the $n$-qubit protocol is superior. We provide a technique that can alleviate certain types of noise. We show that for certain types of noisy evolution the $n$-qubit protocol will be inferior but for others it will be superior. 
\end{abstract}

%\pacs{03.65.Ta, 03.67.-a,03.65.Ud}

\maketitle

%%%%%%%%%%%%%%%%%%%%%%%%%%%%%%%%%%%%%%%%%%%%%%%%%%%%%%%%%%%%%%%%%%%%%%%%%%%%%%%%%%%%%%%%%%%%%%%%%%%%%%%%%
%%%%%%%%%%%%%%%%%%%                                                                       %%%%%%%%%%%%%%%
%%%%%%%%%%%%%%%%%%%         Begin section                                                 %%%%%%%%%%%%%%%
%%%%%%%%%%%%%%%%%%%                                                                       %%%%%%%%%%%%%%%
%%%%%%%%%%%%%%%%%%%%%%%%%%%%%%%%%%%%%%%%%%%%%%%%%%%%%%%%%%%%%%%%%%%%%%%%%%%%%%%%%%%%%%%%%%%%%%%%%%%%%%%%%

\section{Introduction}
\label{sec:intro}

Quantum metrology considers using quantum systems to determine parameters that describe physical processes or states. Metrology takes the view that parameter determination must be described in terms of particular physical systems, specific operations and measurements on those systems and subsequent processing of measurement outcomes. All but the outcome processing are subject to the physical laws that govern the particular systems and these will constrain possible parameter determination procedures and partly dictate their effectiveness. 

Quantum metrology provides a framework for analyzing any scheme for determining a physical parameter. This combines classical statistics and quantum theory to quantify the effectiveness of schemes and to decide which may be preferable~\cite{helstrom76,caves81,braunstein92,braunstein94,lee02,sarovar06,giovannetti04,giovannetti06,paris09, kok10, giovannetti11,kolodynski13,toth14,demkowicz15,pezze18,braun18}. This has frequently been applied to estimate parameters that govern particular qubit evolutions. In many cases the effectiveness of an estimation process using manifestly quantum resources such as entangled states can be greater than that attained by comparable ``classical'' processes~\cite{giovannetti06,fujiwara01,sasaki02,frey11,fujiwara03,ji08,fujiwara04}. 

Studies in quantum metrology initially considered absolutely optimal schemes, using pure initial quantum states. However, some systems, such as solution-state nuclear magnetic resonance (NMR), cannot access pure states and this raises the question of what advantages exist when only mixed states are available~\cite{dariano05, datta11,modi11,pinel13,collins13, collins15,micadei15,serrano25}. Results here were unified for a large class of qubit evolution processes, described by a single parameter to be estimated, on systems with highly mixed initial states~\cite{collins19}. A collection of $n$ qubits, each in the same highly mixed state, are initially subjected to a parameter-independent preparatory unitary process and subsequently only one of the qubits is subjected to the parameter-dependent evolution process; the rest act as spectators. For very highly mixed states, this yields a roughly $n$-fold estimation accuracy gain versus a single qubit process~\cite{collins19}. 

In~\cite{collins19}, the spectator qubits do not evolve after the preparatory unitary. The present article addresses how the presence of additional, typically noisy, evolution on the spectators affects estimation. For example, consider estimating the dephasing time for a single nuclear spin within a multi-spin molecule~\cite{slichter96}. The additional nuclear spins could be used as spectators to enhance the estimation accuracy according to the protocol of~\cite{collins19}. However, they would surely suffer their own dephasing while that of the spin of interest is being probed. How might this affect the enhancement promised in~\cite{collins19}? Are there ways to reduce the effects of spectator dephasing?  

Noisy initial state metrology has also been investigated in optical systems~\cite{micadei15, piera21}. This article covers all physical implementations. 

We will provide a method for quantifying the effects of a large class of spectator noise on the estimation accuracy previously obtained~\cite{collins19}. We describe when the additional spectator noise is catastrophic and when it can be tolerated. It will also provide a process of manipulating the spectator qubits so as to reduce the adverse effects of some types of spectator noise. Although the techniques offered here are similar those of~\cite{collins19}, the inclusion of additional spectator noise results in significant modifications to the existing results on noisy parameter estimation and approaches much more realistic scenarios than those analyzed previously~\cite{modi11, collins13, collins15,collins19}.

This article is organized as follows. Section~\ref{sec:qubitchannelmetrology} describes the framework for qubit channel metrology.  Section~\ref{sec:qubitchannel} provides the general description of qubit channels used in the later analysis. Sec.~\ref{sec:noisystatemetrology} describes metrology with highly mixed initial states. Sections~\ref{sec:sqsc} and~\ref{sec:corrstate} describe the two competing estimation protocols considered here and contain the main results that drive the rest of the article. Section~\ref{sec:twisted} describes a technique that can alleviate the detrimental effects of some noise. Section~\ref{sec:measure} describes the final measurements that various protocols could use. Section~\ref{sec:summary} summarizes the estimation procedures and Sec.~\ref{sec:examples} applies this to various examples.

%%%%%%%%%%%%%%%%%%%%%%%%%%%%%%%%%%%%%%%%%%%%%%%%%%%%%%%%%%%%%%%%%%%%%%%%%%%%%%%%%%%%%%%%%%%%%%%%%%%%%%%%%
%%%%%%%%%%%%%%%%%%%                                                                       %%%%%%%%%%%%%%%
%%%%%%%%%%%%%%%%%%%         End section                                                 %%%%%%%%%%%%%%%
%%%%%%%%%%%%%%%%%%%                                                                       %%%%%%%%%%%%%%%
%%%%%%%%%%%%%%%%%%%%%%%%%%%%%%%%%%%%%%%%%%%%%%%%%%%%%%%%%%%%%%%%%%%%%%%%%%%%%%%%%%%%%%%%%%%%%%%%%%%%%%%%%

%%%%%%%%%%%%%%%%%%%%%%%%%%%%%%%%%%%%%%%%%%%%%%%%%%%%%%%%%%%%%%%%%%%%%%%%%%%%%%%%%%%%%%%%%%%%%%%%%%%%%%%%%
%%%%%%%%%%%%%%%%%%%                                                                       %%%%%%%%%%%%%%%
%%%%%%%%%%%%%%%%%%%         Begin section                                                 %%%%%%%%%%%%%%%
%%%%%%%%%%%%%%%%%%%                                                                       %%%%%%%%%%%%%%%
%%%%%%%%%%%%%%%%%%%%%%%%%%%%%%%%%%%%%%%%%%%%%%%%%%%%%%%%%%%%%%%%%%%%%%%%%%%%%%%%%%%%%%%%%%%%%%%%%%%%%%%%%

\section{Qubit channel metrology}
%\label{sec:}
\label{sec:qubitchannelmetrology}

\begin{figure}%
%\fbox{
 %\includegraphics[bb = 80 550 410 720,scale=0.65]{figprotocols.eps}
 \includegraphics[bb = 80 550 410 720,scale=0.65]{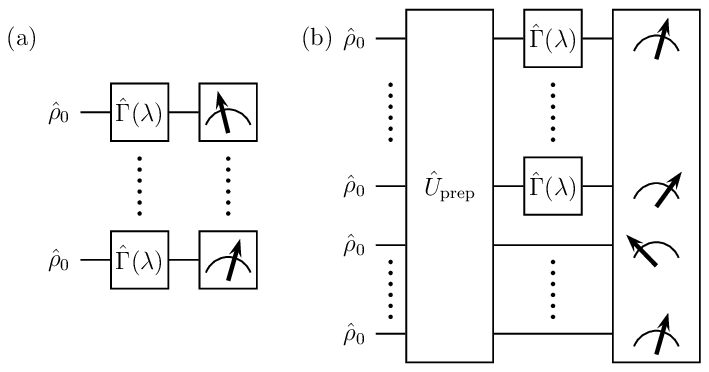}
%}
\caption{Possible metrology protocols. The protocol in a) uses $\channellambda$ independently on each of $q$ qubits. The protocol in b) uses $n$ qubits with $\channellambda$ acting on each of $q$ channel qubits. There are additional spectator qubits, on which the channel does not act. All are subjected to a parameter-independent unitary operation, $\uprep$, prior to querying the channel and a measurement, possibly global, after channel queries.   
				\label{fig:protocols}%
				}
\end{figure}

Suppose that a channel of known type $\channellambda$, but dependent on an unknown parameter, $\lambda$, acts on a single qubit. Metrology considers how to determine the parameter as accurately as possible via physical use of the channel. A protocol to determine $\lambda$ specifies the number of qubits used, their initial states, possible supplementary evolutions, eventual measurements and processing of measurement outcomes. Figure~\ref{fig:protocols} illustrates examples of typical protocols~\cite{dariano01,giovannetti06,ji08,giovannetti11,modi11,kolodynski13,demkowicz14,sekatski17,polino20,len22,jiao23,nielsen23,zhang24}.
Quantum metrology assesses the accuracy of protocols in terms of the statistical fluctuations in parameter estimates and compares competing protocols, aiming for the most accurate.   

A well-established measure of accuracy is the quantum Cram\'{e}r-Rao bound, which bounds the variance in the estimate~\cite{paris09}. Let $m_1,m_2,\ldots$, be the measurement outcomes of the protocol and let $p(m_1, m_2 \ldots|\lambda)$ be the associated probability. An estimate for the parameter is generated via $\lambdaest = \lambdaest(m_1,m_2,\ldots)$, where $\lambdaest(\ldots)$ is an estimator function. The requirements are that the estimator is unbiased, i.e.\ $\expv{\lambdaest} = \lambda$ where $\expv{\lambdaest} = \sum_{m_1, m_2,\ldots} \lambdaest(m_1,m_2,\ldots) p(m_1, m_2 \ldots|\lambda)$ and that the variance $\variance{(\lambdaest)}:= \left< \left( \lambdaest - \expv{\lambdaest} \right)^2 \right>$ is minimized. The variance depends on the choice of estimator but, regardless of this choice, is bounded by classical Cram\'{e}r-Rao bound (CRB)~\cite{cramer46,paris09}, $\variance{(\lambdaest)} \geqslant 1/F(\lambda)$ where the Fisher information is
\[
	F(\lambda) = \sum_{m_1, m_2,\ldots} \left[\frac{\partial \ln{p(m_1, m_2 \ldots|\lambda)}}{\partial \lambda}\right]^2 p(m_1, m_2 \ldots|\lambda). 
\]
The maximum likelihood estimator, determined by solving $ \frac{\partial p }{\partial \lambda}= 0 $ for $\lambda$, saturates the bound~\cite{cramer46} as $n \rightarrow \infty.$

For measurements performed on quantum systems, the system state immediately prior to measurement, $\rhof,$ and the measurement choice dictate the measurement outcome statistics. Regardless of the measurement choice, the Fisher information satisfies the quantum Cram\'{e}r-Rao bound (QCRB) $F(\lambda) \leqslant H(\lambda)$ where the quantum Fisher information (QFI) is determined~\cite{braunstein94,paris09,oloan10,nagaoka12ch9,liu20,polino20} from the pre-measurement state via
\begin{equation}
	H(\lambda) = \Trace{\left[ \rhof(\lambda) \score^2(\lambda)\right]}.
	\label{eq:quantumfisher}
\end{equation}
Here $\score(\lambda)$ is the symmetric logarithmic derivative (SLD) defined implicitly via
\begin{equation}
	\frac{\partial \rhof(\lambda)}{\partial \lambda} = \frac{1}{2}\;
	                                                   \left[
	                                                     \score(\lambda)
	                                                     \rhof(\lambda)
	                                                     +
	                                                     \rhof(\lambda)
	                                                     \score(\lambda)
	                                                   \right].
  \label{eq:slddefintion}
\end{equation}
Thus, regardless of the measurement and estimator choices, $\variance{(\lambdaest)} \geqslant 1/H(\lambda)$. A larger QFI implies a possibly more accurate protocol and thus the QFI can quantify the protocol's effectiveness. 

The QFI can be computed in various ways~\cite{paris09,collins13,kolodynski13,chapeau15,fiderer19}. A QRCB-saturating measurement always exists but may depend on the parameter to be estimated~\cite{braunstein94,paris09,nagaoka12ch9,toscano17}. Here the estimation protocol will be local and adaptive methods can assist~\cite{barndorff00,sekatski17,pang17,rodriguezgarcia21}. 

The remaining choices in any protocol are the number of qubits used, the number and arrangement of channel queries and the input state (i.e.\ the state immediately prior to querying the channel). These will determine $\rhof$ and eventually the QFI. We assume that the only relevant cost for any protocol is the number of channel queries. Then the QFI per channel query is the crucial measure of performance of the protocol, as has often appeared in the literature~\cite{braunstein94,fujiwara01,fujiwara03,fujiwara04,hotta05,sarovar06,giovannetti06,ji08,oloan10,datta11,giovannetti11,modi11,frey11,escher11,zwierz12,kolodynski13,collins13,demkowicz14,collins15,sekatski17,collins19,polino20,len22,jiao23,nielsen23,zhang24}.   

One can compare protocols where each qubit is independent of the others, such as in Fig.~\ref{fig:protocols}(a), to those where the channels act on a correlated or entangled input state, such as in Fig.~\ref{fig:protocols}(b). For certain channels, the latter case can enhance the QFI per channel query~\cite{fujiwara01,fujiwara04,giovannetti06,frey11,polino20}. 

A major focus in quantum metrology has been to determine when entanglement or correlation is advantageous and to quantify this advantage. Such issues are often addressed by optimal resources such as access to pure input states, all possible measurements or noiseless environments. When such resources are absent, the relative advantages of various protocols may be different to those using all possible optimal resources~\cite{escher11,kolodynski13,demkowicz17,sekatski17,collins19,zhou21}.

This work addresses the issue of relative advantages of various protocols, assuming that pure states are not available, studied previously in~\cite{modi11,collins13,collins15,collins19}, with the new additional feature of including unwanted, possibly noisy, in some of the qubits.  

\subsection{Noisy initial states}

We assume that the qubit states at the outset of the protocol (initial states) are noisy with no possibility of noise reduction. A prototypical example is room-temperature, solution-state NMR, which exposes an ensemble of identical molecules to physical interactions which depend on potentially interesting parameters. Nuclear spins in a single molecule provide the qubits used in the protocol; the entire ensemble duplicates the qubits without interaction between the molecules. At room temperature, the state of any spin is highly mixed and there is no practical way to purify it~\cite{slichter96,nielsen00,jones11}. Any metrology protocol on such systems needs to be restricted to noisy initial states and it could be that the relative strengths and weaknesses of competing protocols may differ from the case where pure states are available.  

\begin{figure}%
%\fbox{
 %\includegraphics[bb = 80 580 380 720,scale=0.70]{fignoiseprotocols.eps}
 \includegraphics[bb = 80 580 380 720,scale=0.75]{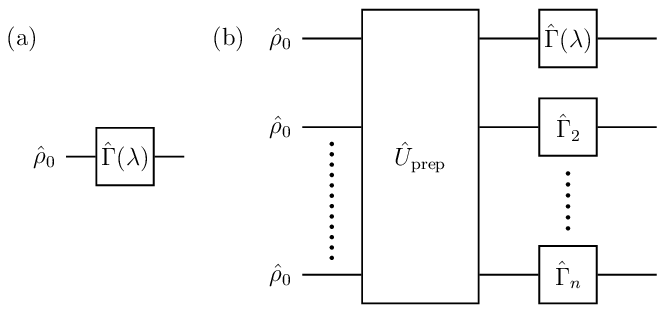}
%}
\caption{Single channel qubit estimation protocols. (a) The SQSC protocol with one qubit and one channel query.  (b) The CS protocol queries $\channellambda$ once on a single qubit with the remaining spectator qubits assisting.  $\channel_2,\ldots,\channel_n$ indicate unwanted but unavoidable noisy channels on the spectators. 
				\label{fig:noiseprotocols}%
				}
\end{figure}

We aim to compare the protocols illustrated in Fig.~\ref{fig:noiseprotocols}.
The \emph{single-qubit, single-channel (SQSC) protocol} uses a single qubit and and single channel query. Its competitor, the \emph{correlated-state (CS) protocol}, uses $n$ qubits, with the channel acting once on one of them and the remaining spectator qubits, possibly assisting, via an input state that potentially correlates the qubits. 

The single parameter-bearing channel use of the CS-protocol is exemplified by determining the dephasing time $T_2$ of one particular spin in each molecule; this is related to the parameter associated with a phase-flip channel~\cite{jones11}. These are typically different for different nuclei in a molecule (e.g.\ alanine~\cite{collins00,teklemariam01,weinstein04} or crotonic acid~\cite{ju10,luxin16,wen19}). If we only wanted to estimate the parameter associated with one nucleus, then this would provide $\channellambda$ while the remaining nuclei would provide spectators. This type of situation been considered previously~\cite{collins13,collins19}. However, if there is dephasing in one spin, it is likely that the spectator spins will also suffer dephasing, that does not depend on $\lambda$. Such more realistic and more important situations have not been considered previously and the central goals of this work are to determine how these affect the accuracy of the CS protocol and whether there are any ways to reduce the effects of such additional noise.   

We assume the same initial states for the protocols of Fig.~\ref{fig:noiseprotocols}, i.e.\ $\rhoo = (\iop + r \myvector{r}_0\cdot \sigmavec)/2$, where $\rinitial$, the \emph{Bloch-sphere initial direction}, is a unit vector in three spatial dimensions and $r$ is the \emph{purity}, satisfying $0 \leqslant r \leqslant 1$. Here $\myvector{r}_0 \cdot \sigmavec := r_{0x}\sigmax + r_{0y}\sigmay + r_{0z}\sigmaz$ (Sometimes we use the notation $\sigmagen{a}  := a_{x}\sigmax + a_{y}\sigmay + a_{z}\sigmaz$ for any vector $\myvector{a}$). We assume that $\rinitial$ can be chosen freely from all possible unit vectors but that the purity cannot be adjusted. 

In the SQSC protocol we assume no further evolution after querying the channel, thus giving a final pre-measurement state $\rhof$ determined via $\rhoi \xmapsto{\channellambda} \rhof$.

In the CS protocol, the channel query is preceded by a preparatory unitary, $\uprep$, that is independent of the parameter. This produces a channel input state $\rhoi= \uprep\rhoo^{\otimes n} \uprep^\dagger.$ Thereafter  $\channellambda$ is queried on one qubit while the remaining qubits are subjected to other channels, assumed to be independent of $\lambda.$  In the CS protocol, the qubits are labeled $1,2,\ldots n$, with qubit~1 being the channel on which $\channellambda$ acts (uppermost in diagrams and leftmost in tensor products) and the remainder the spectator qubits. Denote the channel acting on spectator $j$ by $\channel_j.$ The state after the action of all qubit channels, $\rhof$, is determined via $\rhoi \xmapsto{\channellambda \otimes \channel_2 \cdots \otimes \channel_n} \rhof$.

The remaining assumption is that the initial state is very noisy. In general, if $\channellambda$ is unital (i.e.\ it maps $\iop$ to $\iop$) and without any post-preparation spectator evolution, then for very low purity ($r \ll 1/\sqrt{n}$) the CS protocol always yields a roughly $n$-fold gain in QFI versus the SQSC protocol~\cite{collins19} . This applies to solution-state NMR, where the initial states are thermal equilibrium states are highly mixed~\cite{nielsen00,jones11} with $r \approx 10^{-4}$.

However, that study~\cite{collins19} did not account for non-trivial spectator evolution after $\uprep.$ We remove the $\channel_j = \iop$ constraint and consider possible QFI enhancement in such more realistic scenarios, exemplified by solution-state NMR, than considered previously~\cite{modi11, collins13, collins15,collins19}.

%%%%%%%%%%%%%%%%%%%%%%%%%%%%%%%%%%%%%%%%%%%%%%%%%%%%%%%%%%%%%%%%%%%%%%%%%%%%%%%%%%%%%%%%%%%%%%%%%%%%%%%%%
%%%%%%%%%%%%%%%%%%%                                                                       %%%%%%%%%%%%%%%
%%%%%%%%%%%%%%%%%%%         Begin section                                                 %%%%%%%%%%%%%%%
%%%%%%%%%%%%%%%%%%%                                                                       %%%%%%%%%%%%%%%
%%%%%%%%%%%%%%%%%%%%%%%%%%%%%%%%%%%%%%%%%%%%%%%%%%%%%%%%%%%%%%%%%%%%%%%%%%%%%%%%%%%%%%%%%%%%%%%%%%%%%%%%%

\section{Qubit channel formalism}
%\label{sec:}
\label{sec:qubitchannel}

We briefly digress to discuss a description, essential throughout this work, of qubit channels in terms of mappings of Bloch-sphere vectors.

The state of any qubit is described by a density operator
%
%\begin{equation}
$
	\hat{\rho} = ( 
		\iop + \sigmagen{r}
	)/2
$
%\label{eq:initialstate}
%\end{equation}
%
where $\myvector{r}$ is a vector in three dimensions (the Bloch-sphere vector) with $0 \leqslant \left| \myvector{r} \right| \leqslant 1.$ 

\subsection{Qubit channel evolution}

The evolution of a quantum system is described by a completely-positive, trace-preserving map~\cite{wilde2013} and the associated channel, $\channel,$ 
\begin{equation}
	\rhoi = \frac{1}{2}\, \left( \iop + \rinitiali \cdot \sigmavec \right)
	%\stackrel{\channel}{\mapsto}
	\xmapsto{\channel}
	\rhof = \frac{1}{2}\, \left( \iop + \rfinal \cdot \sigmavec \right)
\end{equation}
where $\rinitiali$ and $\rfinal$ are vectors related by~\cite{nielsen00,chapeau18}
\begin{equation}
	\rfinal = M\rinitiali + \myvector{d}.
\label{eq:blochspheremap}
\end{equation}   
$M$ is a linear operator represented by a $3\times 3$ real matrix, called the \emph{channel Bloch-sphere matrix} and $\myvector{d}$ is a vector. We shall consider unital channels, for which $\iop \stackrel{\channel}{\mapsto} \iop$; here $\myvector{d}=0.$ Thus the channel Bloch-sphere matrix completely characterizes the associated unital channel. 

In the three-dimensional Cartesian orthonormal basis $\left\{ \myvector{x}, \myvector{y}, \myvector{z} \right\}$ the channel Bloch-sphere matrix has form 
\begin{equation}
	M	=	\begin{pmatrix}
				M_{xx} & M_{xy} & M_{xz} \\
				M_{yx} & M_{yy} & M_{yz} \\
				M_{zx} & M_{zy} & M_{zz}
			\end{pmatrix}
%\label{eq:}
\end{equation}
with components, for unital cases,  determined by
\begin{subequations}
\begin{eqnarray}
	M_{xx} & = &	\frac{1}{2}
							\Trace{
								\left[ 
									\sigmax \channel(\sigmax)
								\right]
							},
					\\ 
	M_{xy} & = & 	\frac{1}{2}
							\Trace{
								\left[ 
									\sigmax \channel(\sigmay)
								\right]
							}, \textrm{etc,\ldots}
%\label{eq:}
\end{eqnarray}
\end{subequations}

For example, the unitary phase-shift channel $\rhoi \stackrel{\channel}{\mapsto} \hat{U}^\dagger\rhoi \hat{U}$ where $\hat{U}:= e^{-i\lambda \sigmaz/2}$ is described by
\begin{equation}
	\renewcommand{\arraystretch}{1.25}
	M	=	\begin{pmatrix}
				\cos{\lambda} & -\sin{\lambda} & 0 \\
				\sin{\lambda} & \cos{\lambda} & 0 \\
				0 & 0 & 1
			\end{pmatrix}.
\label{eq:bmuintary}
\end{equation}
Further examples are provided in section~\ref{sec:examples}.

In the metrology protocols that follow, the channel Bloch-sphere matrix for $\channellambda$ will be denoted $M_1$ and those for the spectator channels $\channel_j$ by $M_j$, the latter each assumed to be independent of $\lambda$. 

\subsection{Channel Bloch-sphere matrix products}

We will see that metrology protocols can be assessed in terms of $\dot{M}_1^\top \dot{M}_1$ and $ M_2^\top M_2, \ldots, M_n^\top M_n$ where the dot indicates differentiation with respect to $\lambda.$ We briefly describe their important features. These are positive semi-definite since $\myvector{v}^\top M_j^\top M_j \myvector{v}$, for any vector $\myvector{v}$, is the inner product of $M_j \myvector{v}$ with itself and this is necessarily positive semi-definite. The singular value decomposition for $\dot{M}_1$ and each of $M_j$ implies that $\dot{M}_1^\top \dot{M}_1$ and $ M_2^\top M_2, \ldots, M_n^\top M_n$ can each be diagonalized; the eigenvalues will be positive semi-definite. Thus 
\begin{subequations}
\begin{eqnarray}
	\dot{M}_1^\top \dot{M}_1 	& = \alpha\, \myvector{a}\myvector{a}^\top
															+\beta\, \myvector{b}\myvector{b}^\top
															+\delta\, \myvector{d}\myvector{d}^\top
															\quad
															\textrm{and}
\label{eq:blochspherematrixprodschannel}
															\\
	M_j^\top M_j 							& = \sigma_j \myvector{s}_j\myvector{s}_j^\top
															+\tau_j \myvector{t}_j\myvector{t}_j^\top
															+\upsilon_j \myvector{u}_j\myvector{u}_j^\top
\label{eq:blochspherematrixprods}
\end{eqnarray}
\end{subequations}
where products such as $\myvector{e}\myvector{f}^\top$ signify operators on vectors, defined via $\myvector{e}\myvector{f}^\top \left( \myvector{v} \right): = \myvector{e} \left(\myvector{f} \cdot \myvector{v} \right)$ for any vectors $\myvector{e},\myvector{f},$ and $\myvector{v}.$ The eigenvalues are arranged to satisfy 
$ \alpha \geqslant \beta \geqslant \delta \geqslant 0$ 
and $\left\{ \myvector{a}, \myvector{b}, \myvector{d}\right\}$ form an orthonormal set of vectors, ordered in terms of magnitude of associated eigenvalues. In general, all of these eigenvectors and eigenvalues associated with the channel $\channellambda$ could depend on $\lambda.$ Similarly, 
$1 \geqslant \sigma_j  \geqslant \tau_j \geqslant \upsilon_j  \geqslant 0$ 
and $\left\{ \myvector{s}_j, \myvector{t}_j, \myvector{u}_j \right\}$ form an orthonormal set of vectors, also ordered, but these eigenvectors and eigenvalues are all independent of $\lambda$. 

Note that this description differs from that of~\cite{collins19}, which used singular values of $\dot{M}_1, M_2, \ldots M_n.$ The two descriptions can be reconciled by noting that the eigenvalues of the product matrices are the squares of the singular values of the original matrices.

%%%%%%%%%%%%%%%%%%%%%%%%%%%%%%%%%%%%%%%%%%%%%%%%%%%%%%%%%%%%%%%%%%%%%%%%%%%%%%%%%%%%%%%%%%%%%%%%%%%%%%%%%
%%%%%%%%%%%%%%%%%%%                                                                       %%%%%%%%%%%%%%%
%%%%%%%%%%%%%%%%%%%         Begin section                                                 %%%%%%%%%%%%%%%
%%%%%%%%%%%%%%%%%%%                                                                       %%%%%%%%%%%%%%%
%%%%%%%%%%%%%%%%%%%%%%%%%%%%%%%%%%%%%%%%%%%%%%%%%%%%%%%%%%%%%%%%%%%%%%%%%%%%%%%%%%%%%%%%%%%%%%%%%%%%%%%%%

\section{Metrology with very noisy initial states}
%\label{sec:}
\label{sec:noisystatemetrology}

For the very low purity situations it is advantageous~\cite{collins19} to consider series expansions in $r$.  

The initial state for the system is an $n$-fold tensor product of $\rhoo = (\iop + r\rinitial\cdot\sigmavec)/2$, giving a series in $r$, 
\begin{equation}
	\rhoo^{\otimes n} = \sum_{j=0}^n r^j \rhooorder{j}
%\label{eq:}
\end{equation}
where $\rhooorder{j}$ is independent of $r$. Thus the system input state prior to querying the channel can be expressed as
\begin{equation}
	\rhoi = \sum_{j=0}^n r^j \rhoiorder{j}
%\label{eq:}
\end{equation}
where $\rhoiorder{j} = \uprep \rhooorder{j} \uprep^\dagger$ is also independent of $r$. The final state of the system after querying the channel and prior to measurement can be expressed as
\begin{equation}
	\rhof = \sum_{j=0}^n r^j \rhoforder{j}
%\label{eq:}
\end{equation}
where $\rhoforder{j}$ is independent of $r$ and is determined by action of the channels on $\rhoiorder{j}.$

Similarly the QFI can be expressed as
\begin{equation}
	H = \sum_{j=0}^\infty r^j \Horder{j}
%\label{eq:}
\end{equation}
where $\Horder{j}$ is independent of $r.$

For a \emph{unital channel} it can be shown~\cite{collins19} that 
\begin{subequations}
	\begin{eqnarray}
		\Horder{0} & = & 0,\\
		\Horder{1} & = & 0, \quad \textrm{and} \\
		\Horder{2} & = & N \Trace{\left[ \left( \frac{\partial \rhoforder{1}}{\partial \lambda} \right)^2\right] }.
	\end{eqnarray}
	\label{eq:unitalqfi}
\end{subequations}
where $N=2^n$.

If $r$ is sufficiently small, as described later, then the QFI can be terminated at second order and thus $H \approx r^2 \Horder{2}$. The subsequent analysis follows from Eqs.~\eqref{eq:unitalqfi}.

%%%%%%%%%%%%%%%%%%%%%%%%%%%%%%%%%%%%%%%%%%%%%%%%%%%%%%%%%%%%%%%%%%%%%%%%%%%%%%%%%%%%%%%%%%%%%%%%%%%%%%%%%
%%%%%%%%%%%%%%%%%%%                                                                       %%%%%%%%%%%%%%%
%%%%%%%%%%%%%%%%%%%         Begin section                                                 %%%%%%%%%%%%%%%
%%%%%%%%%%%%%%%%%%%                                                                       %%%%%%%%%%%%%%%
%%%%%%%%%%%%%%%%%%%%%%%%%%%%%%%%%%%%%%%%%%%%%%%%%%%%%%%%%%%%%%%%%%%%%%%%%%%%%%%%%%%%%%%%%%%%%%%%%%%%%%%%%

\section{Single-qubit, single-channel protocols}
\label{sec:sqsc}

We establish a baseline for estimation via the SQSC protocol. The channel input state can be assumed to be the same as the initial state. Thus the estimation protocol is determined by chosing $\rinitial$ and the final post-channel measurement. 

Then $\rhoiorder{0} = \iop$ and $\rhoiorder{1} = \rinitial\cdot\sigmavec,$ giving
\begin{equation}
	\rhoforder{1} = \left( 
										M_1 \rinitial
									\right)
									\cdot
									\sigmavec
%\label{eq:}
\end{equation}
Equation~\eqref{eq:unitalqfi} gives that the lowest-order, non-zero term in the QFI is
\begin{equation}
	H = 	r^2 
				\rinitial^\top
				\dot{M}_1^\top \dot{M}_1
				\rinitial.
\label{eq:sqscqfiordertwo}
\end{equation}
The diagonal form for $\dot{M}_1^\top \dot{M}_1$ given by Eq.~\eqref{eq:blochspherematrixprodschannel} gives~\cite{collins19}
\begin{equation}
	H \leqslant	r^2 \alpha
\label{eq:sqscqfiordertwobound}
\end{equation}
with the upper bound attained for the choice $\rinitial = \myvector{a}.$ 

Thus the \emph{optimal SQSC protocol} uses $\rinitial = \myvector{a}$ and gives lowest optimal order SQSC protocol QFI
\begin{equation}
	\Hsopt = r^2 \alpha.
%\label{eq:}
\end{equation}
Other protocols will be compared to this via their QFIs. 

\subsection{Global versus local protocols}
%\newpage

The optimal SQSC protocol uses $\myvector{a}$, the eigenvector associated with the maximum eigenvalue of $\dot{M}_1^\top \dot{M}_1$. If $\myvector{a}$ is independent of $\lambda$ then the protocol, up to the measurement stage, is constructed globally (independently of the parameter).  Examples of such global estimation protocols are unitary, flip and depolarizing channel parameter estimation (see section~\ref{sec:examples}). 

If $\myvector{a}$ depends on $\lambda$, the optimal protocol requires knowledge of the parameter it aims to determine. Instead a local optimal protocol is possible. This assumes that $\lambda$ is in the vicinity of a known value $\lambda_0$. Let $\myvector{a}_0$ be the known eigenvector associated with the maximum eigenvalue of $\dot{M}_1^\top \dot{M}_1$ when $\lambda = \lambda_0.$ Then (see Appendix~\ref{app:local}), 
\begin{equation}
	\myvector{a}_0^\top
	\dot{M}_1^\top \dot{M}_1
	\myvector{a}_0
	= 
	\myvector{a}^\top
	\dot{M}_1^\top \dot{M}_1
	\myvector{a}
	+ {\cal O}(\lambda - \lambda_0)^2.
\label{eq:sqsclocal}
\end{equation}
Thus, to ${\cal O}(\lambda - \lambda_0)^2$, the optimal QFI is returned using a protocol with $\rinitial = \myvector{a}_0$, corresponding to known parameter value $\lambda_0$. This is a parameter-independent local estimation protocol in the vicinity of~$\lambda_0$.

%%%%%%%%%%%%%%%%%%%%%%%%%%%%%%%%%%%%%%%%%%%%%%%%%%%%%%%%%%%%%%%%%%%%%%%%%%%%%%%%%%%%%%%%%%%%%%%%%%%%%%%%%
%%%%%%%%%%%%%%%%%%%                                                                       %%%%%%%%%%%%%%%
%%%%%%%%%%%%%%%%%%%         Begin section                                                 %%%%%%%%%%%%%%%
%%%%%%%%%%%%%%%%%%%                                                                       %%%%%%%%%%%%%%%
%%%%%%%%%%%%%%%%%%%%%%%%%%%%%%%%%%%%%%%%%%%%%%%%%%%%%%%%%%%%%%%%%%%%%%%%%%%%%%%%%%%%%%%%%%%%%%%%%%%%%%%%%

\section{Symmetric Pairwise Correlated Protocols}
\label{sec:corrstate}

\begin{figure}[h]%
%\fbox{
 %\includegraphics[bb = 70 500 354 720,scale=0.65]{figuprep.eps}
 \includegraphics[bb = 70 500 354 720,scale=0.60]{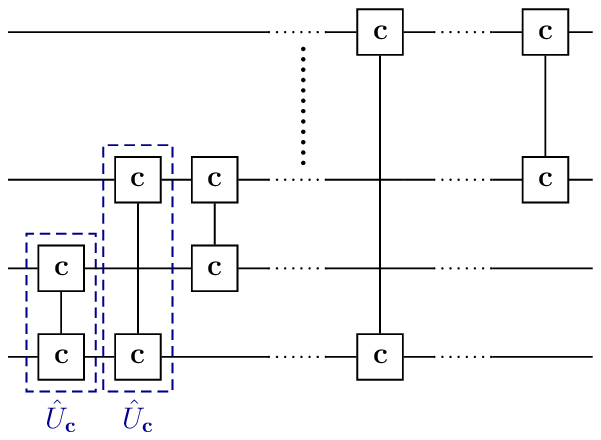}
%}
\caption{Symmetric pairwise correlated protocol preparatory unitary. The circuit element within the blue dashed box a single iteration of $\uc$; the solid boxes indicate the two qubits on which the gate acts.  
				\label{fig:corrscheme}%
				}
\end{figure}

Consider correlating the channel input states via the protocol of Fig.~\ref{fig:noiseprotocols}(b) using a suitable preparatory unitary, $\uprep.$ This can enhance estimation accuracy in various situations when the spectators are noiseless~\cite{collins13,collins15,collins19}. The known results assume a particular preparatory unitary, whose basic component is the two-qubit unitary,
\begin{equation}
  \uc := \frac{1}{2} \left( 
	               \iop \otimes \iop + 
	               \iop \otimes \sigmagen{c} +
								 \sigmagen{c} \otimes \iop - 
								 \sigmagen{c} \otimes \sigmagen{c}
								 \right), 
 %\label{eq:}
\end{equation}
where $\myvector{c}$ is a unit vector, the \emph{Bloch-sphere control direction}. This is applied once to each distinct pair of qubits as illustrated in Fig.~\ref{fig:corrscheme} and the entire collection of such unitaries forms $\uprep,$ which defines the symmetric pairwise correlated protocol. 

The protocol continues with one channel query on one qubit (the channel qubit). For \emph{unital channels with no spectator evolution after the preparatory unitary} this protocol yields~\cite{collins19} a CS-protocol QFI, $\Hcorr$, satisfying
\begin{equation}
	\left( 
		\frac{\beta}{\alpha} + n - 1
	\right) 
	\Hsopt 
	\leqslant \Hcorr \leqslant n \Hsopt,
	\label{eq:qficorronebound}
\end{equation}
where $\alpha$ and $\beta$ are the two largest eigenvalues of $\dot{M}_1^\top \dot{M}_1$ (reference~\cite{collins19} phrases this via singular values). This indicates a roughly $n$-fold gain in the QFI.

Equation~\eqref{eq:qficorronebound} will not necessarily apply when the spectators evolve, for example under environmental noise, after $\uprep$ as illustrated in Fig.~\ref{fig:noiseprotocols}(b). This more realistic situation demands a version of Eq.~\eqref{eq:qficorronebound} that applies when noise acts on the spectators while the channel acts on the channel qubit. We now substantially modify the analysis of~\cite{collins19} to produce a more general version of Eq.~\eqref{eq:qficorronebound}.

The state for the entire system immediately after the preparatory unitary is, to lowest order in the purity, $\rhoi = \rhoiorder{0} + r \rhoiorder{1}$. Using the techniques of~\cite{collins19}, outlined in Appendix~\ref{app:spcinput}, gives  
$
\rhoforder{0} = \iop^{\otimes n}/N
%\label{eq:corrorderzeroinput}
$
and 
\begin{widetext}
\begin{eqnarray}
 \rhoforder{1} 
		& = &
		\frac{1}{N}\;
								\Bigl[
									( M_1 \rinitial )
									\cdot
									\hat{\myvector{\sigma}}
									\otimes
									( M_2 \myvector{c} )
									\cdot
									\hat{\myvector{\sigma}}
									\otimes
									\cdots
									\otimes
									( M_{n} \myvector{c} )
									\cdot
									\hat{\myvector{\sigma}}
								\Bigr. 
		\nonumber \\
		&  & 
		\phantom{\frac{1}{N}\;}
									+
									( M_1 \myvector{c} )
									\cdot
									\hat{\myvector{\sigma}}
									\otimes
									( M_2 \rinitial )
									\cdot
									\hat{\myvector{\sigma}}
									\otimes
									( M_3 \myvector{c} )
									\cdot
									\hat{\myvector{\sigma}}
									\otimes
									\cdots
									\otimes
									( M_{n} \myvector{c} )
									\cdot
									\hat{\myvector{\sigma}}
									+ \cdots +
		\nonumber \\
		&  & 
		\phantom{\frac{1}{N}\;}
									+
									\Bigl.
									( M_1 \myvector{c} )
									\cdot
									\hat{\myvector{\sigma}}
									\otimes
									( M_2 \myvector{c} )
									\cdot
									\hat{\myvector{\sigma}}
									\otimes
									\cdots
									\otimes
									( M_{n-1} \myvector{c} )
									\cdot
									\hat{\myvector{\sigma}}
									\otimes
									( M_{n} \rinitial )
									\cdot
									\hat{\myvector{\sigma}}
									\Bigr]
		\nonumber \\
		&  &
		+
		\frac{\rinitial\cdot{\myvector{c}}}{N}\;
				( M_1 \myvector{c} )
				\cdot
				\hat{\myvector{\sigma}}
				\otimes
				I^{\otimes (n-1)}
		\nonumber \\
		&  &
		-
		\frac{n \rinitial\cdot{\myvector{c}}}{N}\;
				(M_1 \myvector{c} )
				\cdot
				\hat{\myvector{\sigma}}
				\otimes
				( M_2 \myvector{c} )
				\cdot
				\hat{\myvector{\sigma}}
				\otimes
				( M_3 \myvector{c} )
				\cdot
				\hat{\myvector{\sigma}}
				\otimes
				\cdots
				\otimes
				( M_{n} \myvector{c} )
				\cdot
				\hat{\myvector{\sigma}}.
\label{eq:rhoffirstorderunital}
\end{eqnarray}
\end{widetext}
where $M_1(\lambda)$ is the Bloch-sphere matrix for $\channellambda$ and $M_j$ is the Bloch-sphere matrix for channel $\channel_j$.

Assuming \emph{the initial states and preparatory unitary are independent of $\lambda$}, this parameter only appears in $M_1$ and, starting with $\rhoforder{1}$ of Eq.~\eqref{eq:rhoffirstorderunital}, Eqs.~\eqref{eq:unitalqfi} give (see Appendix~\ref{app:qfigen} for details) the lowest order CS QFI as
\begin{eqnarray}
	\Hcorr & = & 
		r^2
			\myvector{V}^\top
				\left(
					\dot{M}_1^\top \dot{M}_1 \otimes M_2^\top M_2 \otimes  \cdots \otimes M_n^\top M_n
				\right)
			\myvector{V}
		\nonumber 
		\\
		& 	& 
		+ r^2
		\left( \rinitial \cdot \myvector{c} \right)^2 \myvector{c}^\top \dot{M}_1^\top \dot{M}_1 \myvector{c}
\label{eq:qfisecond}
\end{eqnarray}
where
\begin{eqnarray}
	\myvector{V} 	& := &  
									\rinitialperp \otimes \myvector{c} \otimes \cdots \otimes \myvector{c}
								+ \myvector{c} \otimes \rinitialperp \otimes \myvector{c} \otimes \cdots \otimes \myvector{c} 
								\nonumber \\
								&  	& 
								+ \cdots +
								\myvector{c} \otimes \cdots \otimes  \myvector{c} \otimes \rinitialperp
\label{eq:qfisecondvector}
\end{eqnarray}
and $\rinitialperp := \rinitial - (\rinitial \cdot \myvector{c}) \myvector{c}$ is the component of $\rinitial$ perpendicular to $\myvector{c}$. In Eqs.~\eqref{eq:qfisecond} and~\eqref{eq:qfisecondvector} the tensor product is that for real vectors. 

Equation~\eqref{eq:qfisecond} establishes the general result for all subsequent analysis. Assume that the type of channel $\channellambda$, but not the parameter value, is known and all information about the spectator channels is known. Then all Bloch sphere matrices can be determined. The task then becomes to find $\rinitial$ and $\myvector{c}$ in Eq.~\eqref{eq:qfisecond} that produce gains in the QFI versus the SQSC protocol or those which maximize the QFI. The situations considered previously~\cite{collins19} are attained by setting $M_j = I$ for $j=2,\ldots, n$ of Eq.~\eqref{eq:qfisecond}, which is therefore a significant generalization. 

\subsection{General QFI bounds}

Finding $\rinitial$ and $\myvector{c}$ that yield the optimal QFI in various situations is not always clear. However, we establish general bounds for the QFI with noisy spectators (see Appendix~\ref{app:qfibounds}). If the eigenvalues of $ M_j^\top M_j$ are $1 \geqslant \sigma_j  \geqslant \tau_j \geqslant \upsilon_j  \geqslant 0$ then, \emph{to lowest order in the purity,}
\begin{equation}
	\Hcorr
	\leqslant
		\begin{cases}
			n \Hsopt\, \Sigma & \text{if $n \geqslant 1/\Sigma $,} \\
			\Hsopt\, & \text{if $n \leqslant 1/\Sigma $} \\
		\end{cases}
\label{eq:qficorrupperbound}
\end{equation}
where $\Sigma = \sigma_2 \ldots \sigma_n$ describes the least destructive effects of spectator noise. Thus for sufficiently destructive spectator noise and low enough numbers of spectators, i.e.\ $n \Sigma  \leqslant 1$, the SQSC protocol provides a better QFI than the CS protocol. This provides an absolute threshold of spectator noise, beyond which the CS protocol is disadvantageous regardless of the choices of $\rinitial$ and $\myvector{c}$. 
 
Similarly (see Appendix~\ref{app:qfibounds}) \emph{to lowest order in the purity,}
\begin{equation}
	\Hcorr
	\geqslant
		\begin{cases}
			n \dfrac{\delta}{\alpha}\, \Hsopt\, \Upsilon & \text{if $n \geqslant 1/\Upsilon$,} \\
			\dfrac{\delta}{\alpha}\, \Hsopt & \text{if $n \leqslant 1/\Upsilon$} \\
		\end{cases}
\label{eq:qficorrlowerbound}
\end{equation}
where $\Upsilon = \upsilon_2 \ldots \upsilon_n$ describes the most destructive effects of spectator noise. However, we shall show that the lower bound can be increased by manipulation of spectator channels and particular choices of Bloch-sphere initial- and control directions. We illustrate this for two extreme cases.

%%%%%%%%%%%%%%%%%%%%%%%%%%%%%%%%%%%%%%%%%%%%%%%%%%%%%%%%%%%%%%%%%%%%%%%%%%%%%%%%%%%%%%%%%%%%%%%%%%%%%%%%%
%%%%%%%%%%%%%%%%%%%                                                                       %%%%%%%%%%%%%%%
%%%%%%%%%%%%%%%%%%%         Begin section                                                 %%%%%%%%%%%%%%%
%%%%%%%%%%%%%%%%%%%                                                                       %%%%%%%%%%%%%%%
%%%%%%%%%%%%%%%%%%%%%%%%%%%%%%%%%%%%%%%%%%%%%%%%%%%%%%%%%%%%%%%%%%%%%%%%%%%%%%%%%%%%%%%%%%%%%%%%%%%%%%%%%

\subsection{Clean-spectator case}

The \emph{clean-spectator case} is that where there is no further evolution of the spectator qubits after the preparatory unitary and before measurement. This is the case considered previously~\cite{collins19}. Then $M_j = I$ for $j=2,\ldots, n$. Here $\sigma_1 = \ldots =\sigma_n =1$ and $\Sigma=1$. In Eq.~\eqref{eq:qficorrupperbound} $n \geqslant 1/\Sigma$ and the lowest order QFI satisfies $\Hcorr \leqslant n \Hsopt.$

For the lower bound, $\upsilon_1 = \ldots =\upsilon_n =1$. Thus $\Upsilon=1$ and $n \Upsilon \geqslant 1$. This would appear to yield a lower bound $\Hcorr \geqslant n \delta/\alpha\, \Hsopt.$ However, the special choices $\rinitial = \myvector{b}$ and $\myvector{c} = \myvector{a},$ where $\myvector{a}$ and $\myvector{b}$ are the ``maximal'' (associated with the two largest eigenvalues) eigenvectors of $\dot{M}_1^\top \dot{M}_1$, and Eq.~\eqref{eq:qfisecond} give
\begin{eqnarray}
	\Hcorr	& = &
							r^2
							\left[
								\beta +
								(n-1) \alpha
							\right]
							\nonumber \\
					& = & \Hsopt
							\left(
								\frac{\beta}{\alpha}
								+ n-1
							\right)
%\label{eq:qficorrcleanboundprelim}
\end{eqnarray}
and straightforward algebra shows that this is never smaller than $n \Hsopt \delta/\alpha$ provided that $n \geqslant 1$ and $\beta \geqslant \delta.$ Since these are assumed, this establishes bounds on the clean spectator CS QFI, $\Hclean$, as
\begin{equation}
	\left(
		\frac{\beta}{\alpha}
		+ n-1
	\right)
	\Hsopt
	\leqslant 
	\Hclean
	\leqslant  
	n 
	\Hsopt,
\label{eq:qficorrcleanbounds}
\end{equation}
which is tighter than Eq.~\eqref{eq:qficorrlowerbound}.
This is equivalent to the expression of~\cite{collins19}. Note that the lower bound is attained by the protocol with $\rinitial = \myvector{b}$ and $\myvector{c} = \myvector{a}$. The upper bound would require a protocol with $\rinitial$ perpendicular to $\myvector{c}$ but there is no guarantee that it can be attained. If  $\myvector{a}$ and $\myvector{b}$ are independent of $\lambda$ then the protocol will be global. If either depends on $\lambda$ then, as in the SQSC case, this could give a local protocol about a known parameter value $\lambda_0$ and this would be correct to ${\cal O}(\lambda-\lambda_0)^2.$

%%%%%%%%%%%%%%%%%%%%%%%%%%%%%%%%%%%%%%%%%%%%%%%%%%%%%%%%%%%%%%%%%%%%%%%%%%%%%%%%%%%%%%%%%%%%%%%%%%%%%%%%%
%%%%%%%%%%%%%%%%%%%                                                                       %%%%%%%%%%%%%%%
%%%%%%%%%%%%%%%%%%%         Begin section                                                 %%%%%%%%%%%%%%%
%%%%%%%%%%%%%%%%%%%                                                                       %%%%%%%%%%%%%%%
%%%%%%%%%%%%%%%%%%%%%%%%%%%%%%%%%%%%%%%%%%%%%%%%%%%%%%%%%%%%%%%%%%%%%%%%%%%%%%%%%%%%%%%%%%%%%%%%%%%%%%%%%

\subsection{Depolarizing-spectator case}

Suppose that each spectator is subjected to a depolarizing channel after the preparatory unitary and before measurement. This maps $\rhoi \xmapsto{\channel_j} \epsilon_j \Trace{[\rhoi]}\iop/2 + (1-\epsilon_j)\rhoi$ where $0 \leqslant \epsilon_j \leqslant 1$ is the depolarizing channel parameter~\cite{wilde2013} (a lower parameter implies weaker noise). Then $M_j = (1-\epsilon_j) \iop$ and $M_j^\top M_j = \sigma_j I$ where $\sigma_j = (1-\epsilon_j)^2.$  Equation~\eqref{eq:qfisecond} gives
\begin{eqnarray}
	\Hcorr
	& = & 
		r^2
			\myvector{V}^\top
				\left(
					\dot{M}_1^\top \dot{M}_1 \otimes I \otimes  \cdots \otimes I
				\right)
			\myvector{V}
			\,
			\prod_{j=2}^n (1-\epsilon_j)^2
		\nonumber 
		\\
		& 	& 
		+ r^2
		\left( \rinitial \cdot \myvector{c} \right)^2 \myvector{c}^\top \dot{M}_1^\top \dot{M}_1 \myvector{c}
\label{eq:qfiseconddepol}
\end{eqnarray}
When $\rinitial$ and $\myvector{c}$ are perpendicular, this reduces the clean spectator correlated-state QFI by a multiplicative factor. This reduction extends throughout any subsequent analysis. For example, the clean spectator QFI bounds Eq.~\eqref{eq:qficorrcleanbounds} are reduced by a factor of $\prod_{j=2}^n (1-\epsilon_j)^2$.

%%%%%%%%%%%%%%%%%%%%%%%%%%%%%%%%%%%%%%%%%%%%%%%%%%%%%%%%%%%%%%%%%%%%%%%%%%%%%%%%%%%%%%%%%%%%%%%%%%%%%%%%%
%%%%%%%%%%%%%%%%%%%                                                                       %%%%%%%%%%%%%%%
%%%%%%%%%%%%%%%%%%%         Begin section                                                 %%%%%%%%%%%%%%%
%%%%%%%%%%%%%%%%%%%                                                                       %%%%%%%%%%%%%%%
%%%%%%%%%%%%%%%%%%%%%%%%%%%%%%%%%%%%%%%%%%%%%%%%%%%%%%%%%%%%%%%%%%%%%%%%%%%%%%%%%%%%%%%%%%%%%%%%%%%%%%%%%

\section{Spectator noise alleviation}
\label{sec:twisted}

\begin{figure}%
	%\fbox{
		%\includegraphics[scale=0.90]{figtwisted.eps}
		\includegraphics[scale=0.90]{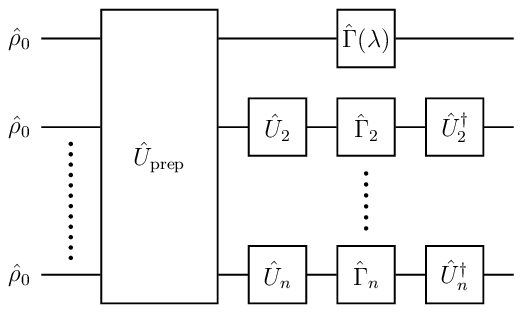}
	%}
	\caption{Symmetric pairwise correlated protocol with twisted spectator channels. Each spectator noisy channel is preceded by a channel-dependent unitary $\hat{U}_j$ and is followed by $\hat{U}_j^\dagger.$
		\label{fig:twisted}%
	}
\end{figure}

One can alleviate the effects of the spectator noise on the estimation accuracy and improve the QFI beyond the lower bound suggested by Eq.~\eqref{eq:qficorrlowerbound}. Inserting appropriately chosen unitaries before and after each spectator channel evolution, as illustrated in Fig.~\ref{fig:twisted}, can modify these so as to reduce their destructive effects.

The strategy that raised the lower bound for the clean-spectator case, resulting in Eq.~\eqref{eq:qficorrcleanbounds} is suggestive. That used the two (``maximal'') eigenvectors associated with the largest eigenvalues of $\dot{M}_1^\top \dot{M}_1$. In Eq.\eqref{eq:qfisecond} the contributions to the QFI from the spectator channels will be determined by the same two maximal eigenvectors and each of $M_j^\top M_j$. It would be advantageous for these eigenvectors to extract the largest eigenvalues of each of $M_j^\top M_j$. The strategy is then to manipulate the spectator channels so as to align their maximal eigenvectors with those of $\dot{M}_1^\top \dot{M}_1$. 

For example, suppose that $\channel_j$ is a phase-flip channel, i.e.\ $\rhoi \xmapsto{\channel_j} (1-\lambda_j) \rhoi + \lambda_j \sigmaz \rhoi \sigmaz$ where $0\leqslant \lambda_j \leqslant 0$. If $\hat{U}_j = e^{i \sigmay \pi/4}$, then, under the combination $\hat{U}_j \rightarrow \channel_j \rightarrow \hat{U}_j^\dagger$,
\begin{equation}
	\rhoi	\mapsto (1-\lambda_j) \rhoi + \lambda_j \sigmax \rhoi \sigmax.
	%\label{eq:}
\end{equation}  
Thus the particular unitary has transformed the phase-flip channel into a bit-flip channel with the same parameter~$\lambda_j$. We call this process \emph{twisting the channel.}

This has important effects on the associated Bloch-sphere matrices. In the example the Bloch-sphere matrix for the phase-flip is
\begin{equation}
	M_j =	\begin{pmatrix}
					1-2\lambda_j & 0 & 0 \\
					0 & 1-2\lambda_j & 0 \\
					0 & 0 & 1
				\end{pmatrix}
%\label{eq:}
\end{equation}
while that for the bit flip is
\begin{equation}
	M_j^\prime =	\begin{pmatrix}
					1 & 0 & 0 \\
					0 & 1-2\lambda_j & 0 \\
					0 & 0 & 1-2\lambda_j
				\end{pmatrix}.
%\label{eq:}
\end{equation}
These are related by the Bloch-sphere matrix for $\hat{U}_j$, 
\begin{equation}
	R_j	=	\begin{pmatrix}
					0 & 0 & -1 \\
					0 & 1 & 0 \\
					1 & 0 & 0
				\end{pmatrix},
%\label{eq:}
\end{equation}
and $M_j^\prime  = R_j^\top	M_j R_j.$ This particular example of twisting would be advantageous if the maximal eigenvectors of $\dot{M}_1^\top \dot{M}_1$ were $\unitvec{x}$ and $\unitvec{y}$. Without twisting these eigenvectors would always extract factors of $1-2\lambda_j$ in Eq.~\eqref{eq:qfisecond} but with the indicated twisting they would sometime extract factors of $1-2\lambda_j$ and other times factors of $1$. The twisted version would give a larger QFI. 

Denote the combination $\hat{U}_j \rightarrow \channel_j \rightarrow \hat{U}_j^\dagger$ by $\channel_j^\prime.$  In general, $\hat{U}_j$ will be represented by an orthogonal rotation Bloch-sphere matrix $R_j$ and the Bloch-sphere matrices for $\channel_j$ and $\channel_j^\prime$ will be related by $M_j^\prime  = R_j^\top	M_j R_j.$ We can always choose $\hat{U}_j$ so as to rotate the eigenvectors of $M_j^\top M_j$ to coincide with the eigenvectors of $\dot{M}_1^\top \dot{M}_1$ in the same decreasing order. The relevant single qubit unitary corresponds to the Bloch-sphere matrix $R_j =  \myvector{s}_j\myvector{a}^\top + \myvector{t}_j\myvector{b}^\top +\myvector{u}_j\myvector{d}^\top.$ Then $M_j^{\prime\top} M_j^\prime = R_j^\top	M_j^\top M_j R_j,$ gives
\begin{equation}
  M_j^{\prime\top} M_j^\prime  = \sigma_j \myvector{a} \myvector{a}^{\top}
																+\tau_j \myvector{b} \myvector{b}^{\top}
																+\upsilon_j \myvector{d} \myvector{d}^{\top}.
%\label{eq:}
\end{equation}
The expression for the QFI of Eq.~\eqref{eq:qfisecond} could then be modified by replacing the Bloch-sphere matrices by their primed versions, \emph{provided that each rotation $R_j$ is independent of $\lambda.$} The rotations are constructed from the eigenvectors of $M_j^{\top} M_j$, which are, by assumption, independent of $\lambda$, and also the eigenvectors of $\dot{M}_1^\top \dot{M}_1$. In some cases (e.g.\ phase shift, Pauli channel, depolarizing channel) the latter are independent of $\lambda$ but there are possibilities where they are not. If the eigenvectors do depend on $\lambda$, the expression of Eq.~\eqref{eq:qfisecond} would need to include derivatives of the spectator channel Bloch-sphere matrices; this will yield a local estimation protocol.    

%%%%%%%%%%%%%%%%%%%%%%%%%%%%%%%%%%%%%%%%%%%%%%%%%%%%%%%%%%%%%%%%%%%%%%%%%%%%%%%%%%%%%%%%%%%%%%%%%%%%%%%%%
%%%%%%%%%%%%%%%%%%%                                                                       %%%%%%%%%%%%%%%
%%%%%%%%%%%%%%%%%%%         Begin section                                                 %%%%%%%%%%%%%%%
%%%%%%%%%%%%%%%%%%%                                                                       %%%%%%%%%%%%%%%
%%%%%%%%%%%%%%%%%%%%%%%%%%%%%%%%%%%%%%%%%%%%%%%%%%%%%%%%%%%%%%%%%%%%%%%%%%%%%%%%%%%%%%%%%%%%%%%%%%%%%%%%%

\subsection{Parameter-independent twisting}

Suppose that the eigenvectors of $\dot{M}_1^\top \dot{M}_1$ are independent of $\lambda.$ Then the spectator channels could be twisted by parameter-independent unitaries and thus take the form (dropping the prime superscripts for convenience)
\begin{equation}
  M_j^{\top} M_j = \sigma_j \myvector{a} \myvector{a}^{\top}
										+\tau_j \myvector{b} \myvector{b}^{\top}
										+\upsilon_j \myvector{d} \myvector{d}^{\top}.
%\label{eq:}
\end{equation}
Choosing $\rinitial=\myvector{b}$ and $\myvector{c}= \myvector{a}$, Eq.~\eqref{eq:qfisecond} gives
\begin{eqnarray}
	\Hcorr	& = &	r^2 \alpha 
					\left(
						\frac{\beta}{\alpha}
						+ \frac{\tau_2}{\sigma_2}
						+ \cdots
						+ \frac{\tau_n}{\sigma_n}
					\right)\, \Sigma 
					\nonumber \\
		& = & \Hsopt\,
					\left(
						\frac{\beta}{\alpha}
						+ \frac{\tau_2}{\sigma_2}
						+ \cdots
						+ \frac{\tau_n}{\sigma_n}
					\right)\, \Sigma .
%\label{eq:}
\end{eqnarray}
Straightforward algebra shows this exceeds $n\Hsopt \Upsilon$ of Eq.~\eqref{eq:qficorrlowerbound}. Thus with twisted spectator channels, 
\begin{equation}
	\Hsopt
	\left(
		\frac{\beta}{\alpha}
		+\sum_{j=2}^n
		  \frac{\tau_j}{\sigma_j}
	\right)
	\Sigma 
	\leqslant \Hcorr \leqslant
	n \Hsopt \Sigma.
\label{eq:twistedbounds}
\end{equation}
Thus when twisting can be done independently of $\lambda,$ a potentially fruitful CS protocol uses twisting and the Bloch-sphere direction choices $\rinitial=\myvector{b}$ and $\myvector{c}= \myvector{a}$.

%%%%%%%%%%%%%%%%%%%%%%%%%%%%%%%%%%%%%%%%%%%%%%%%%%%%%%%%%%%%%%%%%%%%%%%%%%%%%%%%%%%%%%%%%%%%%%%%%%%%%%%%%
%%%%%%%%%%%%%%%%%%%                                                                       %%%%%%%%%%%%%%%
%%%%%%%%%%%%%%%%%%%         Begin section                                                 %%%%%%%%%%%%%%%
%%%%%%%%%%%%%%%%%%%                                                                       %%%%%%%%%%%%%%%
%%%%%%%%%%%%%%%%%%%%%%%%%%%%%%%%%%%%%%%%%%%%%%%%%%%%%%%%%%%%%%%%%%%%%%%%%%%%%%%%%%%%%%%%%%%%%%%%%%%%%%%%%

\subsection{Parameter-dependent twisting}

Suppose that only $\myvector{a}$ depends on $\lambda.$ Then let $\myvector{a}_0$ be the eigenvector associated with the maximum eigenvalue of $\dot{M}_1 ^\top\dot{M}_1$ when the parameter assumes a known value, $\lambda_0.$ Then, using the language of Appendix~\ref{app:local} the twisting rotation is 
\begin{equation}
	R_j  =  \myvector{s}_j\left( \myvector{a}^\top_0 + \dot{\myvector{a}}^\top_0  \Delta\lambda \right)+ \myvector{t}_j\myvector{b}^\top +\myvector{u}_j\myvector{d}^\top + {\cal O}(\Delta\lambda)^2
%\label{eq:}
\end{equation}
where $\Delta\lambda = \lambda-\lambda_0.$ This gives
\begin{eqnarray}
  M_j^{\top} M_j	& =	& \sigma_j \myvector{a}_0 \myvector{a}^{\top}_0
												+\tau_j \myvector{b}_0 \myvector{b}^{\top}_0
												+\upsilon_j \myvector{d}_0 \myvector{d}^{\top}_0
												\nonumber \\
									& 	& + \sigma_j 
													\left( 
														\myvector{a}_0 \dot{\myvector{a}}^{\top}_0 + \dot{\myvector{a}}_0 \myvector{a}^{\top}_0 
													\right)
													\Delta\lambda + {\cal O}(\Delta\lambda)^2. 
%\label{eq:}
\end{eqnarray}
As in Appendix~\ref{app:local}, the orthogonality of $\myvector{a}_0$ and each of $\dot{\myvector{a}}_0, \myvector{b}_0$ and $\myvector{c}_0$ gives that using a protocol where $\rinitial=\myvector{b}_0$ and $\myvector{c}= \myvector{a}_0$ yields Eq.~\eqref{eq:twistedbounds} correct to ${\cal O}(\Delta\lambda)^2$. The same argument extends to cases where all the eigenvectors of $\dot{M}_1 ^\top\dot{M}_1$ are parameter-dependent. 

To summarize, if the eigenvectors of $\dot{M}_1 ^\top\dot{M}_1$ are parameter-dependent, then the twisting could be invoked using eigenvectors at some known parameter value. The resulting protocol will then return the same bounds as for the parameter-independent case but will only be accurate to second order in the deviation from this known parameter value. This will be a local estimation protocol. 

%%%%%%%%%%%%%%%%%%%%%%%%%%%%%%%%%%%%%%%%%%%%%%%%%%%%%%%%%%%%%%%%%%%%%%%%%%%%%%%%%%%%%%%%%%%%%%%%%%%%%%%%%
%%%%%%%%%%%%%%%%%%%                                                                       %%%%%%%%%%%%%%%
%%%%%%%%%%%%%%%%%%%         Begin section                                                 %%%%%%%%%%%%%%%
%%%%%%%%%%%%%%%%%%%                                                                       %%%%%%%%%%%%%%%
%%%%%%%%%%%%%%%%%%%%%%%%%%%%%%%%%%%%%%%%%%%%%%%%%%%%%%%%%%%%%%%%%%%%%%%%%%%%%%%%%%%%%%%%%%%%%%%%%%%%%%%%%

\section{Saturating measurements for symmetric pairwise correlated protocol}
\label{sec:measure}

We ask whether there is a measurement such that the resulting classical Fisher information, $F$, attains the lower bound QFI of Eq.~\eqref{eq:twistedbounds}. It does not appear that there is one measurement scheme that works in all circumstances. Consider the symmetric pairwise correlated protocol with spectators that are twisted and where $\rinitial$ is perpendicular to $\myvector{c}$. We describe two possibilities, one which saturates the bound in certain situations and the other which does not but is close for large $n$. Both possibilities only apply when the the \emph{channel Bloch-sphere matrix for each spectator can be diagonalized}. Some, but not all, examples of this are spectator channels that are unitary, flip-type or depolarizing.  

\subsection{Generic measurement on all qubits}

\begin{figure}%
%\fbox{
 %\includegraphics[bb = 100 600 354 720,scale=0.75]{figgenericmeas.eps}
 %\includegraphics[scale=0.75]{figgenericmeas.eps}
 \includegraphics[scale=0.75]{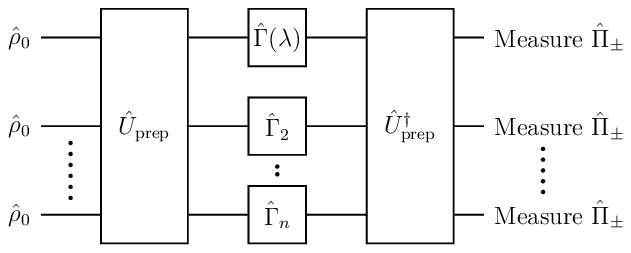}
%}
\caption{Generic measurement scheme. The projection operators associated with the measurement are $\hat{\Pi}_\pm = \left( \iop \pm \rinitial\cdot \sigmavec\right)/2.$  
				\label{fig:genericmeas}%
				}
\end{figure}

A generic scheme invokes the inverse of the preparatory unitary after all channel actions and follows this by single qubit projective measurement along the initial Bloch-sphere direction each qubit, as illustrated in Fig.~\ref{fig:genericmeas}. A projective measurement with operators $\hat{\Pi}_\pm = (\iop \pm  \rinitial\cdot\sigmavec)/2$ is ultimately carried out on each qubit. 

The Fisher information is computed from the probabilities of the measurement outcomes. Assuming the choices $\rinitial = \myvector{b}$ and $\myvector{c} = \myvector{a}$, we show (see Appendix~\ref{app:rhomeasure}) that \emph{to lowest order in the purity}, 
\begin{equation}
 F = r^2 
			\left[
				\left( \rinitial^\top \dot{M}_1 \rinitial \right)^2
				+
				\left( \myvector{c}^\top \dot{M}_1 \myvector{c} \right)^2
					\sum_{j=2}^n
					\frac{\tau_j}{\sigma_j}
			\right]
			\Sigma.
\label{eq:Fgenericmeas}
\end{equation}

This will not always be able to produce the lower bound of Eq.~\eqref{eq:twistedbounds}. For example, it cannot do this if the parameter-dependent channel is a unitary phase shift (see section~\ref{sec:exampleunitary}). 

However, suppose that $\dot{M}_1$ admits a real diagonal decomposition, i.e.\ $\dot{M}_1 = m_1 \myvector{e}_1 \myvector{e}_1^\top + m_2 \myvector{e}_2 \myvector{e}_2^\top + m_3 \myvector{e}_3 \myvector{e}_3^\top$, where $\left\{ \myvector{e}_1, \myvector{e}_2, \myvector{e}_3 \right\}$ form an real orthonormal basis and $m_1, m_2, m_3$ are real. Then the diagonal basis for this is the same as that for $\dot{M}_1^\top\dot{M}_1$ and the eigenvalues of the latter are the squares of the former. Then the choices $\rinitial = \myvector{b}$ and $\myvector{c} = \myvector{a}$ give
\begin{equation}
 F = r^2 
			\left(
				\beta
				+
				\alpha
					\sum_{j=2}^n
					\frac{\tau_j}{\sigma_j}
			\right)
			\Sigma,
%\label{eq:Fgenericmeas}
\end{equation}
and this is exactly the lower bound of Eq.~\eqref{eq:twistedbounds}.

Thus, \emph{if $\dot{M}_1$ admits a real diagonal decomposition,} then choosing $\rinitial = \myvector{b}$ and $\myvector{c} = \myvector{a}$ gives a Fisher information of 
\begin{equation}
 F = \Hsopt
			\left(
				\frac{\beta}{\alpha}
				+
				\sum_{j=2}^n
				\frac{\tau_j}{\sigma_j}
			\right)
			\Sigma.
\label{eq:Fgenericmeasdiag}
\end{equation} 

Note that for clean spectators, $\tau_j = \sigma_j$ for each channel and Eq.~\eqref{eq:Fgenericmeasdiag} gives
\begin{equation}
 F = \Hsopt
			\left(
				\frac{\beta}{\alpha}
				+
				n-1
			\right),
%\label{eq:Fgenericmeasdiag}
\end{equation} 
which is the result of~\cite{collins19}, with $\alpha = s_1^2$ and $\beta= s_2^2,$ giving an almost $n$-fold enhancement of the Fisher information.

%
%%%%%%%%%%%%%%%%%%%%%%%%%%%%%%%%%%%%%%%%%%%%%%%%%%%%%%%%%%%%%%%%%%%%%%%%%%%%%%%%%%%%%%%%%%%%%%%%%%%%%%%%%
%%%%%%%%%%%%%%%%%%%                                                                       %%%%%%%%%%%%%%%
%%%%%%%%%%%%%%%%%%%         Begin section                                                 %%%%%%%%%%%%%%%
%%%%%%%%%%%%%%%%%%%                                                                       %%%%%%%%%%%%%%%
%%%%%%%%%%%%%%%%%%%%%%%%%%%%%%%%%%%%%%%%%%%%%%%%%%%%%%%%%%%%%%%%%%%%%%%%%%%%%%%%%%%%%%%%%%%%%%%%%%%%%%%%%

\subsection{Tailored measurement scheme}

\begin{figure}%
%\fbox{
 %\includegraphics[bb = 100 600 354 720,scale=0.75]{figgenericmeas.eps}
 %\includegraphics[scale=0.75]{figtailoredmeas.eps}
 \includegraphics[scale=0.75]{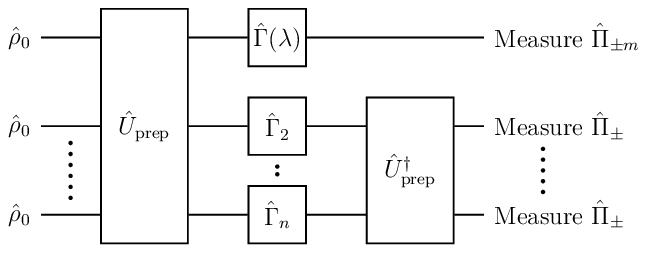}
%}
\caption{Tailored measurement scheme. The measurement operators on the channel qubit are $\hat{\Pi}_{\pm m} = \left( \iop \pm \sigmagen{m}\right)/2.$ Those on the spectators are each $\hat{\Pi}_\pm = \left( \iop \pm \rinitial\cdot\sigmavec\right)/2.$   
				\label{fig:tailoredmeas}%
				}
\end{figure}

If $\dot{M}_1$ does not admit a real diagonal decomposition, consider the modified measurement scheme of Fig.~\ref{fig:tailoredmeas}.

Here the pre-measurement unitary is restricted to the spectator qubits and we allow a general measurement direction on the parameter-dependent qubit. This is described by a unit vector $\unitvec{m}$, the Bloch-sphere measurement direction, and the associated projection operators, $\hat{\Pi}_{\pm m} = \left( \iop \pm \sigmagen{m}\right)/2.$

Then \emph{to lowest order in the purity} (see Appendix~\ref{app:rhomeasurestailored}), 
\begin{equation}
	F	=	r^2
			\left( 
				\myvector{m}^\top \dot{M}_1 \myvector{c} 
			\right)^2
			\left(
				\sum_{j=2}^n
				\frac{\tau_j}{\sigma_j}
			\right)\,
			\Sigma.
\label{eq:Ftailoredmeas}
\end{equation}
The protocol choices require the singular value decomposition, $\dot{M}_1 = ASB$ where $A$ and $B$ are orthogonal matrices and $S$ is diagonal with non-negative entries (the singular values of $\dot{M}_1$). Let $\myvector{e}_1, \myvector{e}_2$ and $\myvector{e}_3$ eigenvectors of $S$ in order of decreasing singular values of $\dot{M}_1$.  Note that $\dot{M}_1^\top \dot{M}_1 = B^\top S^2 B$ reveals that the singular values are $ \sqrt{\alpha} \geqslant \sqrt{\beta}  \geqslant \sqrt{\delta}$ and the eigenvectors of $\dot{M}_1^\top \dot{M}_1$ are $\myvector{a} = B^\top \myvector{e}_1$, $\myvector{b} = B^\top \myvector{e}_2$, and $\myvector{d} = B^\top \myvector{e}_3$. Then choosing $\myvector{c} = B^\top \myvector{e}_1$, $\rinitial = B^\top \myvector{e}_2$ and $\myvector{m} = A \myvector{e}_1$, gives $\left( \myvector{m}^\top \dot{M}_1 \myvector{c} \right)^2 = \alpha$. The explicit choice of Bloch-sphere measurement direction is determined by the details of the parameter-bearing channel and thus the measurement must be \emph{tailored to the channel}. 

Thus there are choices of Bloch-sphere control- and measurement directions so that the tailored measurement scheme yields, \emph{to lowest order in the purity}, 
\begin{equation}
 F = r^2
		\Hsopt
		\left(
			\sum_{j=2}^n
			\frac{\tau_j}{\sigma_j}
		\right)\,
		\Sigma.
\label{eq:Ftailoredmeasbest}
\end{equation}

Note that for clean spectators, $\tau_j = \sigma_j$ for each channel and Eq.~\eqref{eq:Fgenericmeasdiag} gives
\begin{equation}
 F = r^2 \Hsopt (n-1).
%\label{eq:Fgenericmeasdiag}
\end{equation} 
This corrects the result of~\cite{collins19}, which overlooked cases where $\dot{M}_1$ might not admit a real diagonalization. This fails to saturate the lower bound for the QFI, but there is negligible difference if $n$ is large enough. 

Again if any of $A, B$ or $\myvector{e}_1, \myvector{e}_2$ and $\myvector{e}_3$ depend on $\lambda,$ then local protocols using variants of these for a known parameter value $\lambda_0$, will give protocols whose results match those of parameter-independent cases correct to order $(\lambda - \lambda_0)^2.$ 

%%%%%%%%%%%%%%%%%%%%%%%%%%%%%%%%%%%%%%%%%%%%%%%%%%%%%%%%%%%%%%%%%%%%%%%%%%%%%%%%%%%%%%%%%%%%%%%%%%%%%%%%%
%%%%%%%%%%%%%%%%%%%                                                                       %%%%%%%%%%%%%%%
%%%%%%%%%%%%%%%%%%%         Begin section                                                 %%%%%%%%%%%%%%%
%%%%%%%%%%%%%%%%%%%                                                                       %%%%%%%%%%%%%%%
%%%%%%%%%%%%%%%%%%%%%%%%%%%%%%%%%%%%%%%%%%%%%%%%%%%%%%%%%%%%%%%%%%%%%%%%%%%%%%%%%%%%%%%%%%%%%%%%%%%%%%%%%

\section{Summary of the Estimation Protocol}
\label{sec:summary}

The correlated state estimation protocol offered here can be summarized as follows:

\begin{enumerate}
	\item Describe the parameter-dependent channel by the channel Bloch-sphere matrix $M_1$. Describe the spectator channels by channel Bloch-sphere matrices $M_j$ for $j=2,\ldots n.$
	\item Determine the diagonal decompositions
	\begin{eqnarray}
			\dot{M}_1^\top \dot{M}_1 	& = \alpha\, \myvector{a}\myvector{a}^\top
																	+\beta\, \myvector{b}\myvector{b}^\top
																	+\delta\, \myvector{d}\myvector{d}^\top
																	\quad
																	\textrm{and}
		%\label{eq:blochspherematrixprodschannel}
																	\\
			M_j^\top M_j 							& = \sigma_j \myvector{s}_j\myvector{s}_j^\top
																	+\tau_j \myvector{t}_j\myvector{t}_j^\top
																	+\upsilon_j \myvector{u}_j\myvector{u}_j^\top
		%\label{eq:blochspherematrixprods}	
		%\label{eq:}
	\end{eqnarray}
where $\alpha \geqslant \beta \geqslant \delta \geqslant 0$ and $1 \geqslant \sigma_j  \geqslant \tau_j \geqslant \upsilon_j  \geqslant 0.$
	\item Let $\Sigma := \sigma_2 \ldots \sigma_n$. If $n\Sigma \leqslant 1$ then the SQSC protocol is optimal regardless of any other considerations. If $n\Sigma > 1$ then the CS protocol might yield an advantage.
	\item Implement the symmetric pairwise correlated protocol of Fig.~\ref{fig:twisted}, where the $\hat{U}_j$ is the unitary corresponding to Bloch-sphere rotation $R_j =  \myvector{s}_j\myvector{a}^\top + \myvector{t}_j\myvector{b}^\top +\myvector{u}_j\myvector{d}^\top.$ Choose $\myvector{c} =\myvector{a}$ and $\rinitial =\myvector{b}.$ The CS protocol QFI is bounded from below via
	\begin{equation}
		\Hsopt
		\left(
			\frac{\beta}{\alpha}
			+
			\sum_{j=2}^n
			\frac{\tau_j}{\sigma_j}
		\right)
		\Sigma 
		\leqslant \Hcorr.
	\label{eq:figeneric}
	\end{equation}
	\item Determine whether $\dot{M}_1$ admits a real diagonal decomposition. If it does, use the generic measurement scheme of Fig.~\ref{fig:genericmeas} yielding a Fisher information that saturates the lower bound, i.e.\ 
	\begin{equation}
		F= 
		\Hsopt
		\left(
			\frac{\beta}{\alpha}
			+
			\sum_{j=2}^n
			\frac{\tau_j}{\sigma_j}
		\right)
		\Sigma.
	%\label{eq:}
	\end{equation}
	If $\dot{M}_1$ does not admit a real diagonal decomposition, determine the singular value decomposition $\dot{M}_1 = A S B$, where $A$ and $B$ are orthogonal matrices and $S$ is a singular value matrix. Then use the tailored measurement scheme of Fig.~\ref{fig:tailoredmeas} with $\myvector{m} =  A \myvector{e}_1$ where $\myvector{e}_1$ is the vector associated with the maximum singular value in $S.$ This will yield a Fisher information less than the lower bound, i.e.\ 
	\begin{equation}
		F= 
		\Hsopt
		\left(
			\sum_{j=2}^n
			\frac{\tau_j}{\sigma_j}
		\right)\,
		\Sigma.
	%\label{eq:}
	\end{equation}
\end{enumerate}

If $\dot{M}_1^\top \dot{M}_1$ has parameter dependent eigenvectors, $\left\{ \myvector{a},\myvector{b},\myvector{c} \right\},$ then the choices of Bloch-sphere initial-state and control vectors and the twisting unitaries will be parameter independent. A general-parameter dependent protocol is impossible. However if $\lambda$ is known to be close to $\lambda_0$, then a local scheme with protocol choices arising from $\lambda_0$ will allow for an estimation protocol correct to order $(\lambda - \lambda_0)^2.$

Note that the approximations used in the series approach are only useful~\cite{collins19} provided that $nr^2 \ll 1$ and this limits the number of useful spectators in the correlated-state protocol to $n \ll 1/r^2.$

%%%%%%%%%%%%%%%%%%%%%%%%%%%%%%%%%%%%%%%%%%%%%%%%%%%%%%%%%%%%%%%%%%%%%%%%%%%%%%%%%%%%%%%%%%%%%%%%%%%%%%%%%
%%%%%%%%%%%%%%%%%%%                                                                       %%%%%%%%%%%%%%%
%%%%%%%%%%%%%%%%%%%         Begin section                                                 %%%%%%%%%%%%%%%
%%%%%%%%%%%%%%%%%%%                                                                       %%%%%%%%%%%%%%%
%%%%%%%%%%%%%%%%%%%%%%%%%%%%%%%%%%%%%%%%%%%%%%%%%%%%%%%%%%%%%%%%%%%%%%%%%%%%%%%%%%%%%%%%%%%%%%%%%%%%%%%%%

\section{Examples for specific channels}
\label{sec:examples}

We illustrate the process for specific channels.

%%%%%%%%%%%%%%%%%%%%%%%%%%%%%%%%%%%%%%%%%%%%%%%%%%%%%%%%%%%%%%%%%%%%%%%%%%%%%%%%%%%%%%%%%%%%%%%%%%%%%%%%%
%%%%%%%%%%%%%%%%%%%                                                                       %%%%%%%%%%%%%%%
%%%%%%%%%%%%%%%%%%%         Begin section                                                 %%%%%%%%%%%%%%%
%%%%%%%%%%%%%%%%%%%                                                                       %%%%%%%%%%%%%%%
%%%%%%%%%%%%%%%%%%%%%%%%%%%%%%%%%%%%%%%%%%%%%%%%%%%%%%%%%%%%%%%%%%%%%%%%%%%%%%%%%%%%%%%%%%%%%%%%%%%%%%%%%

\subsection{Parameter-dependent unitary channel}
\label{sec:exampleunitary}

Consider the unitary phase shift $\hat{U}:= e^{-i\lambda \sigmaz/2}$. Then
\begin{equation}
	\renewcommand{\arraystretch}{1.25}
	M_1	=	\begin{pmatrix}
				\cos{\lambda} & -\sin{\lambda} & 0 \\
				\sin{\lambda} & \cos{\lambda} & 0 \\
				0 & 0 & 1
			\end{pmatrix}
%\label{eq:bmuintary}
\end{equation}
in the basis $\{ \myvector{x}, \myvector{y}, \myvector{z}\}$ and
\begin{equation}
			\dot{M}_1^\top \dot{M}_1 	= \myvector{x}\myvector{x}^\top
																	+ \myvector{y}\myvector{y}^\top,
%\label{eq:}
\end{equation} 
giving $\alpha=\beta=1.$ The optimal SQSC protocol chooses $\rinitial$ to be any vector in the $xy$ plane and gives $\Hsopt = r^2.$

The CS protocol QFI lower bound is attained with any $\rinitial$ and $\myvector{c}$ in the $xy$-plane and perpendicular to each other (e.g.\ $\myvector{c} = \myvector{x}, \rinitial = \myvector{y}$) and is
\begin{equation}
	\Hsopt
	\left(
		1
		+ 
		\sum_{j=2}^n
		\frac{\tau_j}{\sigma_j}
	\right)
	\Sigma
	\leqslant 
	\Hcorr.
\label{eq:qfilowerunitary}
\end{equation}
Whether this allows for an enhancement in the QFI depends on the spectator noise. Some examples are:

\emph{Spectator channels are unitary:} Here $\tau_j=\sigma_j=1$ and thus $\Sigma=1.$ Then Eqs.~\eqref{eq:qfilowerunitary} and~\eqref{eq:qficorrupperbound} implies that 
\begin{equation}
\Hcorr = n \Hsopt
\end{equation}
and that this is the \emph{optimal CS protocol.} However, although $\dot{M}_1$ is a normal matrix, it has eigenvalues $\pm i e^{\pm i \lambda}$ and eigenvectors $(\myvector{x} \mp i \myvector{y})/2$ and thus does not admit a real diagonal decomposition. The generic measurement scheme yields a Fisher information with terms $(\rinitial^\top \dot{M}_1 \rinitial)^2 = (\myvector{c}^\top \dot{M}_1 \myvector{c})^2 = \sin^2{\lambda}$ and would give $F = n \sin^2{\lambda} \Hsopt.$ The tailored measurement scheme, using Bloch-sphere measurement vector $\myvector{m} = -\sin{\lambda}\, \unitvec{x} + \cos{\lambda}\, \unitvec{y}$ gives $F = (n-1)\Hsopt$. Since the measurement protocol is parameter-dependent, this protocol would only be suitable for local parameter estimation.  

The QFI is the same as for the clean-spectator case. Here the spectator channel unitaries could be absorbed into the measurement without affecting the QFI and with no need for twisting unitaries.  

\emph{Spectator channels are flips:} Here, $\rhoi \xmapsto{\channel_j} (1-\lambda_j) \rhoi + \lambda_j \sigmagen{n} \rhoi \sigmagen{n}$ where $\unitvec{n}$ is any unit vector and $0 \leqslant \lambda_j \leqslant 1$. Twisting could convert this to a bit-flip, for which $\myvector{n} = \myvector{x}$ and
\begin{equation}
	\renewcommand{\arraystretch}{1.25}
	M_j	=	\begin{pmatrix}
				1 & 0 & 0 \\
				0 & 1-2 \lambda_j  & 0 \\
				0 & 0 & 1-2 \lambda_j
			\end{pmatrix}.
%\label{eq:bmuintary}
\end{equation}
Here $\sigma_j=1$ and $\tau_j = (1-2 \lambda_j)^2$. Thus $\Sigma=1.$ Then Eq.~\eqref{eq:qfilowerunitary} implies that 
\begin{equation}
			\Hsopt
			\left( 
				1 + \sum_{j=2}^n \tau_j 
			\right)
			\leqslant
			\Hcorr.
%\label{eq:}
\end{equation}
The CS protocol always yields an enhanced QFI. The measurement choice is again constrained by $\dot{M}_1$. The generic measurement scheme would give $F = \sin^2{\lambda}(1+ \tau_2 + \ldots \tau_n) \Hsopt.$ The tailored measurement scheme, using Bloch-sphere measurement vector $\myvector{m} = -\sin{\lambda}\, \unitvec{x} + \cos{\lambda}\, \unitvec{y}$ gives $F = (\tau_2 + \ldots \tau_n)\Hsopt$. For sufficiently weak spectator noise, $\sum_j \tau_j = \sum_j(1-2 \lambda_j)^2 > 1$ and this would provide an enhancement. This would again only be suitable for local parameter estimation. 

\emph{Spectator channels are depolarizing:} Here, $\rhoi \xmapsto{\channel_j} \epsilon_j\Trace{[\rhoi]} \iop/2 +  (1-\epsilon_j)  \rhoi$ where $0\leqslant \epsilon_j \leqslant 1$ is the depolarizing parameter. Then $M_j = (1-\epsilon_j) \iop$, giving $\sigma_j = \tau_j = \upsilon_j = (1-\epsilon_j)^2.$ Equations~\eqref{eq:qfilowerunitary} and~\eqref{eq:qficorrupperbound} imply 
\begin{equation}
	\Hcorr	=
					n
					\Hsopt
					\Sigma
%\label{eq:}
\end{equation}
where $\Sigma = \prod_j (1-\epsilon_j)^2.$ This is the \emph{optimal CS protocol QFI.} The CS protocol can yield an enhanced QFI if the spectator depolarization is weak enough, i.e.\ $\prod_j (1-\epsilon_j)^2 > 1/n.$ The same tailored measurement scheme used for the previous two cases would yield a Fisher information $F = (n-1) \Hsopt \prod_j (1-\epsilon_j)^2$ and this would be advantageous whenever $\prod_j (1-\epsilon_j)^2 > 1/(n-1).$ 

%%%%%%%%%%%%%%%%%%%%%%%%%%%%%%%%%%%%%%%%%%%%%%%%%%%%%%%%%%%%%%%%%%%%%%%%%%%%%%%%%%%%%%%%%%%%%%%%%%%%%%%%%
%%%%%%%%%%%%%%%%%%%                                                                       %%%%%%%%%%%%%%%
%%%%%%%%%%%%%%%%%%%         Begin section                                                 %%%%%%%%%%%%%%%
%%%%%%%%%%%%%%%%%%%                                                                       %%%%%%%%%%%%%%%
%%%%%%%%%%%%%%%%%%%%%%%%%%%%%%%%%%%%%%%%%%%%%%%%%%%%%%%%%%%%%%%%%%%%%%%%%%%%%%%%%%%%%%%%%%%%%%%%%%%%%%%%%

\subsection{Parameter-dependent flip channel}

The flip channel maps $\rhoi \xmapsto{\channellambda} (1-\lambda) \rhoi + \lambda \sigmagen{n} \rhoi \sigmagen{n}$ where $\unitvec{n}$ is any unit vector. All such channels are related to the phase-flip channel ($\myvector{n} = \myvector{z}$) by a parameter-independent rotation and we can restrict the analysis to this. Here
\begin{equation}
	\renewcommand{\arraystretch}{1.25}
	M_1=	\begin{pmatrix}
				1-2 \lambda  & 0 & 0\\
				0 & 1-2 \lambda & 0 \\
				0 & 0 & 1 
			\end{pmatrix},
%\label{eq:bmuintary}
\end{equation} 
in the basis $\{ \myvector{x}, \myvector{y}, \myvector{z}\}$ and
\begin{equation}
	\renewcommand{\arraystretch}{1.25}
	\dot{M}_1^\top \dot{M}_1 = 4 \myvector{x}\myvector{x}^\top
														+ 4 \myvector{y}\myvector{y}^\top.
\label{eq:bmprodpauli}
\end{equation}
Thus $\alpha = 4$ ($\myvector{a} = \myvector{x}$), $\beta = 4$ ($\myvector{b} = \myvector{y}$) and $\delta=0$ ($\myvector{d} = \myvector{z}$). The optimal SQSC protocol uses a Bloch-sphere initial direction in the $xy$-plane and the optimal QFI is $\Hsopt = 4r^2.$ This protocol is parameter-independent.

The CS protocol QFI lower bound again uses any $\rinitial$ and $\myvector{c}$ in the $xy$-plane and perpendicular to each other (e.g.\ $\myvector{c} = \myvector{x}, \rinitial = \myvector{y}$) and is
\begin{equation}
	\Hsopt
	\left(
		1
		+ \sum_{j=2}^n
		\frac{\tau_j}{\sigma_j}
	\right)
	\Sigma
	\leqslant 
	\Hcorr.
\label{eq:qfilowerflip}
\end{equation}
Examples of some spectator channels are: 

\emph{Spectator channels are unitary:} Here $\tau_j=\sigma_j=1$ and thus $\Sigma=1.$ Equations~\eqref{eq:qfilowerflip} and~\eqref{eq:qficorrupperbound} imply that 
\begin{equation}
\Hcorr = n \Hsopt
\end{equation}
and this is the \emph{optimal CS protocol.} Here $\dot{M}_1$ can be diagonalized independently of the parameter. The generic measurement scheme gives $F = n \Hsopt$. 

\emph{Spectator channels are flips:} The spectator channels can be twisted into bit-flip channels. Then $\sigma_j=1$ and $\tau_j = (1-2 \lambda_j)^2$. Thus $\Sigma=1.$ Then Eq.~\eqref{eq:qfilowerflip} implies that 
\begin{equation}
			\Hsopt
			\left( 
				1 + \sum_{j=2}^n \tau_j 
			\right)
			\leqslant 
			\Hcorr.
%\label{eq:}
\end{equation}
The CS protocol will always yield an enhanced QFI. The generic measurement protocol can be used and, for this, Eq.~\eqref{eq:figeneric} gives a Fisher information 
$F= \Hsopt
			\left( 
				1 + \tau_2 +\cdots +\tau_n
			\right).
$
This saturates the QFI lower bound and yields a Fisher information larger than the SQSC protocol. 

\emph{Spectator channels are depolarizing:} Equations~\eqref{eq:qfilowerflip} and~\eqref{eq:qficorrupperbound} imply
\begin{equation}
	\Hcorr
	=
	n
	\Hsopt
	\Sigma
%\label{eq:}
\end{equation}
where $\Sigma = \prod_j (1-\epsilon_j)^2.$ The generic measurement scheme will yield $F= n \Hsopt \prod_j (1-\epsilon_j)^2$ and the CS protocol plus generic measurement will be advantageous whenever $\prod_j (1-\epsilon_j)^2 > 1/n$. This is also the \emph{optimal CS protocol.}

%%%%%%%%%%%%%%%%%%%%%%%%%%%%%%%%%%%%%%%%%%%%%%%%%%%%%%%%%%%%%%%%%%%%%%%%%%%%%%%%%%%%%%%%%%%%%%%%%%%%%%%%%
%%%%%%%%%%%%%%%%%%%                                                                       %%%%%%%%%%%%%%%
%%%%%%%%%%%%%%%%%%%         Begin section                                                 %%%%%%%%%%%%%%%
%%%%%%%%%%%%%%%%%%%                                                                       %%%%%%%%%%%%%%%
%%%%%%%%%%%%%%%%%%%%%%%%%%%%%%%%%%%%%%%%%%%%%%%%%%%%%%%%%%%%%%%%%%%%%%%%%%%%%%%%%%%%%%%%%%%%%%%%%%%%%%%%%

\subsection{Parameter-dependent depolarizing channel}

Here $\rhoi \xmapsto{\channellambda} \lambda\Trace{[\rhoi]} \iop/2 +  (1-\lambda) \rhoi$. Then 
\begin{equation}
	M_1  = (1-\lambda)  I,
\end{equation}
giving
\begin{equation}
	\dot{M}_1^\top\dot{M}_1  = I.
\end{equation}
Then $\alpha = \beta =\delta=1$, with the eigenvectors any set of three orthogonal vectors. The optimal QSQC protocol uses any vector and gives $\Hsopt = r^2.$ This protocol is parameter independent. 

The correlated-state protocol QFI lower bound again uses any $\rinitial$ and $\myvector{c}$ in the $xy$-plane and perpendicular to each other. Choosing $\myvector{c} = \myvector{x}$ and $\rinitial = \myvector{y}$  gives
\begin{equation}
	\Hsopt
	\left(
		1
		+	\sum_{j=2}^n
		\frac{\tau_j}{\sigma_j}
	\right)
	\Sigma 
	\leqslant
	\Hcorr.
\label{eq:qfilowerdepol}
\end{equation}
Then $\dot{M}_1$ can be diagonalized independently of the parameter and the generic measurement scheme can be used, giving a Fisher information that saturates the QFI lower bound. Further examples are: 

\emph{Spectator channels are unitary:} Then Eq.~\eqref{eq:qfilowerdepol} implies that 
\begin{equation}
\Hcorr = n \Hsopt 
%\label{eq:}
\end{equation}
and the generic measurement scheme gives $F = n \Hsopt$. This is the \emph{optimal CS protocol.} 

\emph{Spectator channels are flips:} Then Eq.~\eqref{eq:qfilowerdepol} implies that 
\begin{equation}
	\Hsopt
	\left( 
		1 + \sum_{j=2}^n \tau_j
	\right)
	\leqslant
	\Hcorr
%\label{eq:}
\end{equation}
where $\tau_j =(1-2 \lambda_j)^2$. The generic measurement protocol can be used and, for this, Eq.~\eqref{eq:figeneric} gives a Fisher information 
$F= \Hsopt
			\left( 
				1 + \tau_2 +\cdots +\tau_n
			\right).
$
This saturates the QFI lower bound and yields a Fisher information larger than the SQSC protocol. 

\emph{Spectator channels are depolarizing:} Equations~\eqref{eq:qfilowerdepol} and~\eqref{eq:qficorrupperbound} imply 
\begin{equation}
	\Hcorr
	=
	n
	\Hsopt
	\Sigma
%\label{eq:}
\end{equation}
where $\Sigma = \prod_j (1-\epsilon_j)^2.$ The generic measurement scheme gives $F= n \Hsopt \prod_j (1-\epsilon_j)^2$ and the CS protocol plus generic measurement will be advantageous whenever $\prod_j (1-\epsilon_j)^2 > 1/n$. This is again the \emph{optimal CS protocol.} 

These examples show that, if the spectator channels are unitary or flip channels, then there are always choices of protocol directions (i.e. $\myvector{c}$ and $\rinitial$) that yield a QFI larger than the SQSC protocol. Similarly, if the spectator channels are depolarizing, then the perpendicular choices of protocol directions are the optimal possible for the CS protocol but this will only yield an enhanced QFI if the spectator noise is sufficiently weak. 

The unitary phase-shift channel examples also illustrate the general fact that this parameter estimation requires a parameter-dependent measurement (at least for $r<1)$ and is therefore a local estimation scheme. In contrast, for flip and depolarizing channel parameter estimation, the generic measurement scheme will always provide a parameter-independent protocol and a Fisher information that saturates the QFI lower bound. 

%%%%%%%%%%%%%%%%%%%%%%%%%%%%%%%%%%%%%%%%%%%%%%%%%%%%%%%%%%%%%%%%%%%%%%%%%%%%%%%%%%%%%%%%%%%%%%%%%%%%%%%%%
%%%%%%%%%%%%%%%%%%%                                                                       %%%%%%%%%%%%%%%
%%%%%%%%%%%%%%%%%%%         Begin section                                                 %%%%%%%%%%%%%%%
%%%%%%%%%%%%%%%%%%%                                                                       %%%%%%%%%%%%%%%
%%%%%%%%%%%%%%%%%%%%%%%%%%%%%%%%%%%%%%%%%%%%%%%%%%%%%%%%%%%%%%%%%%%%%%%%%%%%%%%%%%%%%%%%%%%%%%%%%%%%%%%%%

\section{Conclusion}
\label{sec:discussion}

We have compared two protocols for metrology of single-parameter, unital, qubit-channel evolutions. These are applicable when the available initial states have very low purity and the channel is queried once. They extend previous results~\cite{collins19} by including additional noise on spectator qubits after input state preparation. 

We have provided a procedure for deciding when the additional noise is so severe that the specific correlated-state protocol under consideration reduces the QFI to below that of a single-qubit protocol. We have also identified spectator noise levels that are low enough that the correlated-state protocol can enhance the estimation accuracy; these depend on the nature of the noise in the spectator channels. If, for example, they undergo a flip-type channel, then the correlated-state protocol can always be chosen to enhance the estimation accuracy. if they are depolarizing channels, then possible enhancement depends on the severity of the depolarization. If this can be characterized via Bloch-sphere operations, then we have provided simple algebraic expressions for deciding when enhancement may or may not occur. 

We have shown that the effects of spectator noise on estimation accuracy can be alleviated by twisting the spectator channels, in a manner sympathetic to the parameter-bearing channel. The twisting process adds two unitary operations to each spectator channel; these are decided by a combination of the parameter-bearing channel and the spectator channel. Twisting can turn a potentially disadvantageous spectator noise situation (for example, flip channels) into one where the correlated-state protocol will be advantageous regardless of the severity of the noise in the spectator channels.  

It remains to determine the effects of non-unital spectator noise, such as amplitude damping, on the ability of the correlated-state protocol to enhance estimation accuracy. Mathematically, this has a slightly different description to unital noise and would require some modifications to the analysis offered here although the general framework would still apply. 

Our analysis considers one particular correlated-state protocol and this could be generalized by generalizing $\uprep.$ Whether there are fruitful alternatives or what the optimal preparatory unitary might be are open questions. 

%%%%%%%%%%%%%%%%%%%%%%%%%%%%%%%%%%%%%%%%%%%%%%%%%%%%%%%%%%%%%%%%%%%%%%%%%%%%%%%%%%%%%%%%%%%%%%%%%%%%%%%%%
%%%%%%%%%%%%%%%%%%%                                                                       %%%%%%%%%%%%%%%
%%%%%%%%%%%%%%%%%%%         End section                                                 %%%%%%%%%%%%%%%
%%%%%%%%%%%%%%%%%%%                                                                       %%%%%%%%%%%%%%%
%%%%%%%%%%%%%%%%%%%%%%%%%%%%%%%%%%%%%%%%%%%%%%%%%%%%%%%%%%%%%%%%%%%%%%%%%%%%%%%%%%%%%%%%%%%%%%%%%%%%%%%%%

%\acknowledgments

%%%%%%%%%%%%%%%%%%%%%%%%%%%%%%%%%%%%%%%%%%%%%%%%%%%%%%%%%%%%%%%%%%%%%%%%%%%%%%%%%%%%%%%%%%%%%%%%%%%%%%%%%
%%%%%%%%%%%%%%%%%%%                                                                       %%%%%%%%%%%%%%%
%%%%%%%%%%%%%%%%%%%         End section                                                 %%%%%%%%%%%%%%%
%%%%%%%%%%%%%%%%%%%                                                                       %%%%%%%%%%%%%%%
%%%%%%%%%%%%%%%%%%%%%%%%%%%%%%%%%%%%%%%%%%%%%%%%%%%%%%%%%%%%%%%%%%%%%%%%%%%%%%%%%%%%%%%%%%%%%%%%%%%%%%%%%

\appendix

%%%%%%%%%%%%%%%%%%%%%%%%%%%%%%%%%%%%%%%%%%%%%%%%%%%%%%%%%%%%%%%%%%%%%%%%%%%%%%%%%%%%%%%%%%%%%%%%%%%%%%%%%
%%%%%%%%%%%%%%%%%%%                                                                       %%%%%%%%%%%%%%%
%%%%%%%%%%%%%%%%%%%         Begin section                                                 %%%%%%%%%%%%%%%
%%%%%%%%%%%%%%%%%%%                                                                       %%%%%%%%%%%%%%%
%%%%%%%%%%%%%%%%%%%%%%%%%%%%%%%%%%%%%%%%%%%%%%%%%%%%%%%%%%%%%%%%%%%%%%%%%%%%%%%%%%%%%%%%%%%%%%%%%%%%%%%%%

\section{Local estimation protocols}
\label{app:local}

Suppose that $\myvector{a}_0, \myvector{b}_0$ and $\myvector{c}_0$ are the ordered eigenvectors of $\dot{M}_1 ^\top \dot{M}_1$ when $\lambda = \lambda_0.$ Then $\myvector{a} = \myvector{a}_0 + (\lambda - \lambda_0) \dot{\myvector{a}}_0 + {\cal O}(\lambda - \lambda_0)^2,$ where $\dot{\myvector{a}}_0 = d\myvector{a}/dt$ evaluated at $\lambda_0$ with similar expressions for the other eigenvectors. Thus
\begin{eqnarray}
	\dot{M}_1 ^\top \dot{M}_1 & = & \alpha \myvector{a}\myvector{a}^\top 
																	+ \beta \myvector{b}\myvector{b}^\top 
																	+ \delta \myvector{d}\myvector{d}^\top
																	\nonumber \\
														& = & \alpha \myvector{a}_0\myvector{a}^\top_0
																	+ \beta \myvector{b}_0\myvector{b}^\top_0 
																	+ \delta \myvector{d}_0\myvector{d}^\top_0
																	\nonumber \\
														&  & + \alpha
																	\left(
																		\dot{\myvector{a}}_0\myvector{a}^\top_0
																		+
																		\myvector{a}_0\dot{\myvector{a}}^\top_0
																	\right)
																	(\lambda - \lambda_0)
																	\nonumber \\
														&  & + \beta
																	\left(
																		\dot{\myvector{b}}_0\myvector{b}^\top_0
																		+
																		\myvector{b}_0\dot{\myvector{b}}^\top_0
																	\right)
																	(\lambda - \lambda_0)
																	\nonumber \\
														&  & + \delta
																	\left(
																		\dot{\myvector{d}}_0\myvector{d}^\top_0
																		+
																		\myvector{d}_0\dot{\myvector{d}}^\top_0
																	\right)
																	(\lambda - \lambda_0)
																	\nonumber \\
														&  & + {\cal O}(\lambda - \lambda_0)^2.
%\label{eq:}
\end{eqnarray}
By the orthonormality of eigenvectors, $\myvector{a}_0\myvector{a}^\top_0 \myvector{a}_0 = \myvector{a}_0, $ $\myvector{a}_0\myvector{a}^\top_0 \myvector{b}_0 = 0$, etc.,\ldots. Then $\myvector{a}^\top \myvector{a} = 1$ implies that $\dot{\myvector{a}}^\top \myvector{a} = 0$. Thus $\dot{\myvector{a}}^\top_0 \myvector{a}_0 = 0$. These give
$
	\myvector{a}^\top_0 \dot{M}_1 ^\top \dot{M}_1\myvector{a}_0  = \alpha + {\cal O}(\lambda - \lambda_0)^2.
$
But the orthonormality of $\{ \myvector{a}, \myvector{b}, \myvector{c} \}$ gives
$
	\myvector{a}^\top \dot{M}_1 ^\top \dot{M}_1\myvector{a}  = \alpha.
$
This establishes Eq.~\eqref{eq:sqsclocal}.

%%%%%%%%%%%%%%%%%%%%%%%%%%%%%%%%%%%%%%%%%%%%%%%%%%%%%%%%%%%%%%%%%%%%%%%%%%%%%%%%%%%%%%%%%%%%%%%%%%%%%%%%%
%%%%%%%%%%%%%%%%%%%                                                                       %%%%%%%%%%%%%%%
%%%%%%%%%%%%%%%%%%%         Begin section                                                 %%%%%%%%%%%%%%%
%%%%%%%%%%%%%%%%%%%                                                                       %%%%%%%%%%%%%%%
%%%%%%%%%%%%%%%%%%%%%%%%%%%%%%%%%%%%%%%%%%%%%%%%%%%%%%%%%%%%%%%%%%%%%%%%%%%%%%%%%%%%%%%%%%%%%%%%%%%%%%%%%

\section{Symmetric pairwise correlated protocol input state}
\label{app:spcinput}

The channel input state can be determined by repeatedly using the following results, proved in~\cite{collins19}. If $\myvector{c}$ is a unit vector and $\myvector{a}$ and $\myvector{b}$ are arbitrary vectors then
\begin{eqnarray}
	\uc \iop \otimes \iop \uc^\dagger									& = &	\iop \otimes \iop, 
																													\nonumber \\
	\uc \sigmagen{a} \otimes \iop \uc^\dagger					& = &	\sigmagen{a} \otimes \sigmagen{c}  
																													\nonumber \\
																										& 	& \left( \myvector{a} \cdot \myvector{c} \right)
																													\left(
																														\sigmagen{c} \otimes \iop - \sigmagen{c} \otimes \sigmagen{c}
																													\right),
																													\nonumber \\
	\uc \iop \otimes \sigmagen{b} \uc^\dagger					& = &	\sigmagen{c} \otimes \sigmagen{b}  
																													\nonumber \\
																										& 	& \left( \myvector{b} \cdot \myvector{c} \right)
																													\left(
																														 \iop \otimes \sigmagen{c}- \sigmagen{c} \otimes \sigmagen{c}
																													\right), 
																													\;
																													\textrm{and}
																													\nonumber \\
	\uc \sigmagen{a} \otimes \sigmagen{b} \uc^\dagger	& = &	\left( \myvector{a} \times \myvector{c} \right)
																													\cdot \hat{\myvector{\sigma}} \otimes 
																													\left( \myvector{b} \times \myvector{c} \right)
																													\cdot
																													\hat{\myvector{\sigma}}
																													\nonumber \\
																										& 	& \left( \myvector{a} \cdot \myvector{c} \right)
																													\iop \otimes \sigmagen{b}
																													+
																													\left( \myvector{b} \cdot \myvector{c} \right)
																													\sigmagen{a} \otimes \iop 
																													\nonumber \\
																										& 	& \left( \myvector{a} \cdot \myvector{c} \right)
																													\left( \myvector{b} \cdot \myvector{c} \right)
																													\left(
																															- \iop \otimes \sigmagen{c} 
																															- \sigmagen{c} \otimes \iop  
																													\right.
																													\nonumber \\
																										& 	&	\left.
																															+ \sigmagen{c} \otimes \sigmagen{c}
																													\right).
	\label{eq:ucaction}
\end{eqnarray}
The zero-order term of the initial state is obtained by acting on $\iop^{\otimes n}/N$ with $\uprep.$ This will produce $\rhoiorder{0} = \iop^{\otimes n}/N$.

The first-order term of the initial state is a linear combination of terms of the form $\rinitial\cdot \sigmavec \otimes \iop \otimes \cdots \otimes \iop$. Then Eqs.~\eqref{eq:ucaction} imply that the $\uc$ factors acting on all but the leftmost qubit leave this invariant. Thus we only need to consider $\uc$ factors acting on the leftmost plus each other qubit. Equation~\eqref{eq:ucaction} gives that $\uc$ acting on the two leftmost qubits maps
\begin{eqnarray}
	\rinitial\cdot\sigmavec \otimes \iop^{\otimes (n-1)} & \mapsto & \left[ 
																																			\rinitial \cdot \sigmavec \otimes \sigmagen{c} \otimes \iop^{\otimes (n-2)}
																																			+
																																\right.
																																\nonumber \\
																										&        &	\phantom{[} 
																																\left( \rinitial \cdot \myvector{c} \right)
																																\sigmagen{c} \otimes \iop^{\otimes (n-1)}
																																\nonumber \\
																										&        &	\left.
																																-	\left( \rinitial \cdot \myvector{c} \right)
																																	\sigmagen{c} \otimes \sigmagen{c} \otimes \iop^{\otimes (n-2)}
																																\right].
%\label{eq:}
\end{eqnarray}
Applying $\uc$ to qubits~1 and~3 will only modify the first term, producing $\rinitial \cdot \sigmavec \otimes \sigmagen{c} \otimes \sigmagen{c} \otimes \iop^{\otimes (n-3)} + (\rinitial \cdot \myvector{c})[\sigmagen{c} \otimes \sigmagen{c} \otimes \iop^{\otimes (n-2)} - \sigmagen{c}^{\otimes 3} \otimes \iop^{\otimes (n-3)}].$  Continuing this for all $\uc$ acting on qubit~1 gives
\begin{eqnarray}
	\rinitial \cdot \sigmavec \otimes \iop^{\otimes (n-1)} & \mapsto & \left[ 
																																			\rinitial \cdot \sigmavec \otimes \sigmagen{c}^{\otimes (n-1)} 
																																			+
																																			\left( \rinitial \cdot \myvector{c} \right)
																																			\sigmagen{c} \otimes \iop^{\otimes (n-1)}
																																\right.
																																\nonumber \\
																										&        &	\left.
																																-	\left( \rinitial \cdot \myvector{c} \right)
																																	\sigmagen{c}^{\otimes n}
																																\right].
%\label{eq:}
\end{eqnarray}
Repeating this for all first-order terms of the initial state yields~\cite{collins19}
\begin{eqnarray}
 \rhoiorder{1} & = &
                \frac{1}{N}\;
								\left(
								  \rinitial \cdot \sigmavec \otimes \sigmagen{c}^{\otimes(n-1)}
									+ \cdots + 
									\sigmagen{c}^{\otimes(n-1)} \otimes \rinitial \cdot \sigmavec
								\right) 
								\nonumber \\
							& \phantom{=} & 
							  +
                \frac{\rinitial \cdot \myvector{c}}{N}\;
								\left(
								  \sigmagen{c} \otimes \iop^{\otimes(n-1)}
									+ \cdots + 
								  \iop^{\otimes(n-1)} \otimes \sigmagen{c}
								\right)
								\nonumber \\
							& \phantom{=} & 
							  - 
								\frac{n\, \rinitial \cdot \myvector{c}}{N}\; 
								\sigmagen{c}^{\otimes n}.							 
\label{eq:corrorderoneinput}
\end{eqnarray}
Then channels $\channellambda, \channel_2, \ldots \channel_n$ collectively map
\begin{widetext}
\begin{eqnarray}
  \iop^{\otimes n} 
			& \mapsto &
				\iop^{\otimes n},
				\nonumber
				\\
 \rinitial \cdot \sigmavec \otimes \sigmagen{c}^{\otimes(n-1)} 
			& \mapsto &
				\left( M_1 \rinitial \right)
				\cdot
				\hat{\myvector{\sigma}}
				\otimes
				\left( M_2 \myvector{c} \right)
				\cdot
				\hat{\myvector{\sigma}}
				\otimes
				\cdots
				\otimes
				\left( M_n \myvector{c} \right)
				\cdot
				\hat{\myvector{\sigma}},
				\nonumber
				\\
  \sigmagen{c} \otimes \rinitial \cdot \sigmavec \otimes \sigmagen{c}^{\otimes(n-2)}
			& \mapsto &
				\left( M_1 \myvector{c} \right)
				\cdot
				\hat{\myvector{\sigma}}
				\otimes
				\left( M_2 \rinitial \right)
				\cdot
				\hat{\myvector{\sigma}}
				\otimes
				\left( M_3 \myvector{c} \right)
				\cdot
				\hat{\myvector{\sigma}}
				\otimes
				\cdots
				\otimes
				\left( M_n \myvector{c} \right)
				\cdot
				\hat{\myvector{\sigma}}, \ldots, 
				\nonumber
				\\
  \sigmagen{c} \otimes \iop^{\otimes (n-1)} 
			& \mapsto &
				\left( M_1 \myvector{c} \right)
				\cdot
				\hat{\myvector{\sigma}}
				\otimes
				\iop^{\otimes (n-1)}, \textrm{and} 
				\nonumber
				\\
  \sigmagen{c}^{\otimes n}
			& \mapsto &
				\left( M_1 \myvector{c} \right)
				\cdot
				\hat{\myvector{\sigma}}
				\otimes
				\left( M_2 \myvector{c} \right)
				\cdot
				\hat{\myvector{\sigma}}
				\otimes
				\cdots
				\otimes
				\left( M_n \myvector{c} \right).
%\label{eq:}
\end{eqnarray}
\end{widetext}
This gives Eq.~\eqref{eq:rhoffirstorderunital}.

%%%%%%%%%%%%%%%%%%%%%%%%%%%%%%%%%%%%%%%%%%%%%%%%%%%%%%%%%%%%%%%%%%%%%%%%%%%%%%%%%%%%%%%%%%%%%%%%%%%%%%%%%
%%%%%%%%%%%%%%%%%%%                                                                       %%%%%%%%%%%%%%%
%%%%%%%%%%%%%%%%%%%         Begin section                                                 %%%%%%%%%%%%%%%
%%%%%%%%%%%%%%%%%%%                                                                       %%%%%%%%%%%%%%%
%%%%%%%%%%%%%%%%%%%%%%%%%%%%%%%%%%%%%%%%%%%%%%%%%%%%%%%%%%%%%%%%%%%%%%%%%%%%%%%%%%%%%%%%%%%%%%%%%%%%%%%%%

\section{Lowest order QFI}
\label{app:qfigen}

The key for Eq.~\eqref{eq:qfisecond} is $\Trace{\left[ \sigmagen{a} \sigmagen{b}\right]} = 2 \myvector{a}^\top \myvector{b}$ for any vectors $\myvector{a}$ and $\myvector{b}.$ Then Eq.~\eqref{eq:unitalqfi} requires the square of the derivative of Eq.~\eqref{eq:rhoffirstorderunital}. Assuming $\rinitial$ and $\myvector{c}$ are independent of $\lambda$, $\rhofderivorder{1}$ is the same as $\rhoforder{1}$ with $M_1$ replaced by $\dot{M}_1.$ The resulting expression for  $\rhofderivorder{1}$ has four types of terms. 

The square of the first type, $
								\Bigl[
									( \dot{M}_1 \rinitial )
									\cdot
									\hat{\myvector{\sigma}}
									\otimes
									( M_2 \myvector{c} )
									\cdot
									\hat{\myvector{\sigma}}
									\otimes
									\cdots
									\otimes
									( M_{n} \myvector{c} )
									\cdot
									\hat{\myvector{\sigma}}
								\Bigr]/N, 
$
produces
\begin{eqnarray}
		&	& 
			N
			\Trace{
				\left\{ 	
					\frac{1}{N}^2\;
						\Bigl[
							\left( \dot{M}_1 \rinitial \right)
							\cdot
							\hat{\myvector{\sigma}}
							\otimes
							\left( M_2 \myvector{c} \right)
							\cdot
							\hat{\myvector{\sigma}}
							\otimes
							\cdots
							\otimes
							\left( M_{n} \myvector{c} \right)
							\cdot
							\hat{\myvector{\sigma}}
						\Bigr]^2 
				\right\}
			}
			\nonumber \\
		&  &
		 =
			\frac{1}{N}\; 
			\Trace{ 
				\left\{ 	 \left[ \dot{M}_1 \rinitial \cdot \hat{\myvector{\sigma}} \right]^2 \right\}
			}
			\prod_{j=2}^n 
			\Trace{ 
				\left\{ 	 \left[ M_j \myvector{c} \cdot \hat{\myvector{\sigma}} \right]^2 \right\}
			}
			\nonumber \\
		&  &
		 =
			\frac{1}{N}\; 2^n
			\left( \rinitial ^\top \dot{M}_1^\top \dot{M}_1 \rinitial\right)
			\prod_{j=2}^n
			\left(
				\myvector{c}^\top M_j^\top M_j \myvector{c}
			\right)
			\nonumber \\
		&  &
		 =
			\left( \rinitial ^\top \dot{M}_1^\top \dot{M}_1 \rinitial\right)
			\prod_{j=2}^n
			\left(
				\myvector{c}^\top M_j^\top M_j \myvector{c}
			\right)
			\nonumber \\
		&  &
		 =
			\myvector{v_1}^\top
				\left(
					\dot{M}_1^\top \dot{M}_1 \otimes M_2^\top M_2 \otimes  \cdots \otimes M_n^\top M_n
				\right)
			\myvector{v_1}.
%\label{eq:}
\end{eqnarray}
where
\begin{equation}
	\myvector{v_1} := \rinitial \otimes \myvector{c} \otimes \cdots \otimes \myvector{c}
	%\label{eq:}
\end{equation}
and 
\begin{equation}
	\myvector{v_1}^\top := \rinitial^\top \otimes \myvector{c}^\top \otimes \cdots \otimes \myvector{c}^\top.
	%\label{eq:}
\end{equation}

Similarly the square of the second, $
								\Bigl[
									( \dot{M}_1 \myvector{c} )
									\cdot
									\hat{\myvector{\sigma}}
									\otimes
									( M_2 \rinitial )
									\cdot
									\hat{\myvector{\sigma}}
									\otimes
									\cdots
									\otimes
									( M_{n} \myvector{c} )
									\cdot
									\hat{\myvector{\sigma}}
								\Bigr]/N, 
$
produces
\begin{eqnarray}
		&	& 
			N
			\Trace{
				\left\{ 	
					\frac{1}{N}^2\;
						\Bigl[
							\left( \dot{M}_1 \myvector{c} \right)
							\cdot
							\hat{\myvector{\sigma}}
							\otimes
							\left( M_2 \rinitial \right)
							\cdot
							\hat{\myvector{\sigma}}
							\otimes
							\cdots
							\otimes
							\left( M_{n} \myvector{c} \right)
							\cdot
							\hat{\myvector{\sigma}}
						\Bigr]^2 
				\right\}
			}
			\nonumber \\
		&  &
		 =
			\myvector{v_2}^\top
				\left(
					\dot{M}_1^\top \dot{M}_1 \otimes M_2^\top M_2 \otimes  \cdots \otimes M_n^\top M_n
				\right)
			\myvector{v_2}.
%\label{eq:}
\end{eqnarray}
where
\begin{equation}
	\myvector{v_2} :=  \myvector{c} \otimes \rinitial \otimes \myvector{c} \otimes \cdots \otimes \myvector{c}.
	%\label{eq
\end{equation}

The product of the first and the second produces
$
			\myvector{v_1}^\top
				\left(
					\dot{M}_1^\top \dot{M}_1 \otimes M_2^\top M_2 \otimes  \cdots \otimes M_n^\top M_n
				\right)
			\myvector{v_2},
$
while the second with the first produces
$
			\myvector{v_2}^\top
				\left(
					\dot{M}_1^\top \dot{M}_1 \otimes M_2^\top M_2 \otimes  \cdots \otimes M_n^\top M_n
				\right)
			\myvector{v_1},
$

The square of the third type of term,
$ 	( \rinitial\cdot{\myvector{c}})
		( \dot{M}_1 \myvector{c} )
				\cdot
				\hat{\myvector{\sigma}}
				\otimes
				I^{\otimes (n-1)}
$
produces $\left( \rinitial\cdot{\myvector{c}} \right)^2 \myvector{c}^\top \dot{M}_1^\top \dot{M}_1 \myvector{c}$.

The products of the third type with any others produce $0$ since they all contain factors of traceless Pauli operators.

The square of the fourth type of term, 
$
		n (\rinitial\cdot{\myvector{c}})\,
				(\dot{M}_1 \myvector{c} )
				\cdot
				\hat{\myvector{\sigma}}
				\otimes
				( M_2 \myvector{c} )
				\cdot
				\hat{\myvector{\sigma}}
				\otimes
				( M_3 \myvector{c} )
				\cdot
				\hat{\myvector{\sigma}}
				\otimes
				\cdots
				\otimes
				( M_{n} \myvector{c} )
				\cdot
				\hat{\myvector{\sigma}}/N
$
produces
$
 n^2 (\rinitial\cdot{\myvector{c}})^2 
	\myvector{u}^\top
	\dot{M}_1^\top \dot{M}_1 \otimes M_2^\top M_2 \otimes  \cdots \otimes M_n^\top M_n
	\myvector{u}
$
where
\begin{equation}
 \myvector{u}:= \myvector{c} \otimes \myvector{c} \otimes \cdots \otimes \myvector{c}.
%\label{eq:}
\end{equation}
The product of the first and fourth types produces
$
 -n (\rinitial\cdot{\myvector{c}})
	\myvector{v_1}^\top
	\dot{M}_1^\top \dot{M}_1 \otimes M_2^\top M_2 \otimes  \cdots \otimes M_n^\top M_n
	\myvector{u}
$
while the product of the fourth with the first produces
$
 -n (\rinitial\cdot{\myvector{c}}) 
	\myvector{u}^\top
	\dot{M}_1^\top \dot{M}_1 \otimes M_2^\top M_2 \otimes  \cdots \otimes M_n^\top M_n
	\myvector{v_1}.
$

Combining all of these contributions gives
\begin{eqnarray}
	\Hcorr = r^2 & & 
			\left[ \rinitial \otimes \myvector{c} \otimes \cdots \otimes \myvector{c}
						+ \myvector{c} \otimes \rinitial \otimes \cdots \otimes \myvector{c} + \cdots
			\right.
			\nonumber \\
			& & 
			  \left.
					+ \myvector{c} \otimes \cdots \otimes \myvector{c} \otimes \rinitial  
					- n \left( \rinitial \cdot \myvector{c} \right) \myvector{c} \otimes \cdots \otimes \myvector{c} \otimes \myvector{c}
				\right]^\top
			\nonumber \\
			& & 
				\left(
					\dot{M}_1^\top \dot{M}_1 \otimes M_2^\top M_2 \otimes  \cdots \otimes M_n^\top M_n
				\right)
			\nonumber \\
			& & 
			\left[ \rinitial \otimes \myvector{c} \otimes \cdots \otimes \myvector{c}
						+ \myvector{c} \otimes \rinitial \otimes  \cdots \otimes \myvector{c} + \cdots
			\right.
			\nonumber \\
			& & 
			  \left.
					+ \myvector{c} \otimes \cdots \otimes \myvector{c} \otimes \rinitial  
					- n \left( \rinitial \cdot \myvector{c} \right) \myvector{c} \otimes \cdots \otimes \myvector{c} \otimes \myvector{c}
				\right]
		\nonumber 
		\\
		& 	& 
		+ r^2
		\left( \rinitial \cdot \myvector{c} \right)^2 \myvector{c}^\top \dot{M}_1^\top \dot{M}_1 \myvector{c}.
%\label{eq:}
\end{eqnarray}
Then
\begin{eqnarray}
			& & 
			\left[ \rinitial \otimes \myvector{c} \otimes \cdots \otimes \myvector{c}
						+ \myvector{c} \otimes \rinitial \otimes \cdots \otimes \myvector{c} + \cdots
			\right.
			\nonumber \\
			& & 
			  \left.
					+ \myvector{c} \otimes \cdots \otimes \myvector{c} \otimes \rinitial  
					- n \left( \rinitial \cdot \myvector{c} \right) \myvector{c} \otimes \cdots \otimes \myvector{c} \otimes \myvector{c}
				\right]
			\nonumber \\
		=	& & 
			\left\{ \left[ \rinitial - \left( \rinitial \cdot \myvector{c} \right)\, \myvector{c} \right] \otimes \myvector{c} \otimes \cdots \otimes \myvector{c}
			\right.
			\nonumber \\
			& & 
						+ \myvector{c} \otimes \left[ \rinitial - \left( \rinitial \cdot \myvector{c} \right)\, \myvector{c} \right] \otimes  \cdots \otimes \myvector{c} + \cdots
			\nonumber \\
			& & 
			  \left.
					+ \myvector{c} \otimes \cdots \otimes \myvector{c} \otimes \left[ \rinitial - \left( \rinitial \cdot \myvector{c} \right)\, \myvector{c} \right]
				\right\}.
%\label{eq:}
\end{eqnarray}
This gives Eq.~\eqref{eq:qfisecond}.

%%%%%%%%%%%%%%%%%%%%%%%%%%%%%%%%%%%%%%%%%%%%%%%%%%%%%%%%%%%%%%%%%%%%%%%%%%%%%%%%%%%%%%%%%%%%%%%%%%%%%%%%%
%%%%%%%%%%%%%%%%%%%                                                                       %%%%%%%%%%%%%%%
%%%%%%%%%%%%%%%%%%%         Begin section                                                 %%%%%%%%%%%%%%%
%%%%%%%%%%%%%%%%%%%                                                                       %%%%%%%%%%%%%%%
%%%%%%%%%%%%%%%%%%%%%%%%%%%%%%%%%%%%%%%%%%%%%%%%%%%%%%%%%%%%%%%%%%%%%%%%%%%%%%%%%%%%%%%%%%%%%%%%%%%%%%%%%

\section{Lowest order QFI bounds}
\label{app:qfibounds}

The QFI of Eq.~\eqref{eq:qfisecond} is calculated from combinations of the form $\myvector{U}^\top P \myvector{U}$ where $P$ is a positive semidefinite operator and $\myvector{U}$ a vector. Then $P = \sum_i p_i \myvector{e}_i \myvector{e}_i^\top$ where $p_i\geqslant 0$ gives
\begin{equation}
	\myvector{U}^\top P \myvector{U} 
	= 
	\sum_i p_i 
	\left( 
		\myvector{U}^\top \myvector{e}_i \myvector{e}_i^\top\myvector{U} 
	\right).
%\label{eq:}
\end{equation}
Each term in parentheses is positive and thus
\begin{equation}
	\myvector{U}^\top P \myvector{U} \leqslant p_\textrm{max} \sum_i \left( \myvector{U}^\top \myvector{e}_i \myvector{e}_i^\top\myvector{U} \right).
%\label{eq:}
\end{equation}
where $p_\textrm{max}$ is the maximum eigenvalue. Then, by the completeness of the eigenbasis, 
\begin{equation}
	\sum_i \myvector{U}^\top \myvector{e}_i \myvector{e}_i^\top\myvector{U} = \myvector{U}^\top\myvector{U}.
%\label{eq:}
\end{equation}
Thus
\begin{equation}
	\myvector{U}^\top P \myvector{U} \leqslant p_\textrm{max} \myvector{U}^\top\myvector{U}.
%\label{eq:}
\end{equation}
Similarly
\begin{equation}
	p_\textrm{min} \myvector{U}^\top\myvector{U} \leqslant \myvector{U}^\top P \myvector{U}
%\label{eq:}
\end{equation}
where $p_\textrm{min}$ is the minimum eigenvalue of $P$.

The operator in the term
\begin{equation}
\myvector{V}^\top
	\left(
		\dot{M}_1^\top \dot{M}_1 \otimes M^\top_2 M_2 \otimes  \cdots \otimes M^\top_n M_n
	\right)
\myvector{V}
\label{eq:qficleanfirstterm}
\end{equation}
has maximum eigenvalue $\alpha \Sigma$  where $\Sigma = \sigma_2 \ldots \sigma_n$ and minimum eigenvalue $\delta \Upsilon$ where $\Upsilon = \upsilon_2 \ldots \upsilon_n$. Thus this term is bounded from below by $\delta \Upsilon \myvector{V}^\top\myvector{V}$ and from above by $\alpha \Sigma \myvector{V}^\top\myvector{V}$. Then 
\begin{equation}
	\myvector{V}^\top\myvector{V} = n 
																	\left( 
																		\rinitialperp \cdot \rinitialperp
																	\right) 
																	\left( 
																		\myvector{c} \cdot \myvector{c}
																	\right)^{n-1}. 
%\label{eq:}
\end{equation}
Then $\rinitialperp \cdot \rinitialperp = 1 - \left(\rinitial\cdot \myvector{c}\right)^2$ gives that the term of Eq.~\eqref{eq:qficleanfirstterm} is bounded from below by 
\begin{equation}
 \delta  n \left[1 - \left(\rinitial \cdot \myvector{c}\right)^2\right] \Upsilon
%\label{eq:}
\end{equation}
and from above by 
\begin{equation}
 \alpha n \left[1 - \left(\rinitial \cdot \myvector{c}\right)^2\right] \Sigma
%\label{eq:}
\end{equation}

Similarly the term $\left(\rinitial \cdot \myvector{c}\right)^2 \myvector{c}^\top \dot{M}_1^\top \dot{M}_1 \myvector{c}$ in Eq.~\eqref{eq:qfisecond} is bounded from above by $\alpha \left(\rinitial \cdot \myvector{c}\right)^2$ and from below by $\delta \left(\rinitial \cdot \myvector{c}\right)^2$.

Both terms in Eq.~\eqref{eq:qfisecond} are positive and this gives an upper bound
\begin{eqnarray}
	\Hcorr	& \leqslant &
		r^2 \alpha 
		\left[
			n \Sigma + \left(\rinitial \cdot \myvector{c}\right)^2
			\left(1-n \Sigma \right)
		\right]
		\nonumber \\
		& = & 
		\Hsopt 
		\left[
			n \Sigma + \left(\rinitial \cdot \myvector{c}\right)^2
			\left(1-n \Sigma \right)
		\right]
\label{eq:qficorrupperboundprelim}
\end{eqnarray}
Then if $n \Sigma >1$ this is maximal when $\rinitial$ is perpendicular to $\myvector{c}$ and this particular case yields an upper bound of $n \Hsopt \Sigma.$ On the other hand, if $n \Sigma <1$ this is maximal when $\rinitial$ is parallel to $\myvector{c}$ and this yields an upper bound of $\Hsopt.$

The same argument yields a lower bound of 
\begin{eqnarray}
	\Hcorr	& \geqslant &
		r^2 \delta 
		\left[
			n \Upsilon + \left(\rinitial \cdot \myvector{c}\right)^2
			\left(1-n \Upsilon \right)
		\right]
		\nonumber \\
		& = & 
		\Hsopt
		\frac{\delta}{\alpha}\,
		\left[
			n \Upsilon + \left(\rinitial \cdot \myvector{c}\right)^2
			\left(1-n \Upsilon \right)
		\right].
\label{eq:qficorrcleanboundprelim}%\label{eq:qficorrcleanboundprelim}
\end{eqnarray}
Then if $n \Upsilon >1$, choosing $\rinitial$ is perpendicular to $\myvector{c}$ gives a lower bound of $n \delta/\alpha \Hsopt.$ If $n \Upsilon <1$, choosing $\rinitial$ is parallel to $\myvector{c}$ gives a lower bound of $\delta/\alpha \Hsopt.$

%%%%%%%%%%%%%%%%%%%%%%%%%%%%%%%%%%%%%%%%%%%%%%%%%%%%%%%%%%%%%%%%%%%%%%%%%%%%%%%%%%%%%%%%%%%%%%%%%%%%%%%%%
%%%%%%%%%%%%%%%%%%%                                                                       %%%%%%%%%%%%%%%
%%%%%%%%%%%%%%%%%%%         Begin section                                                 %%%%%%%%%%%%%%%
%%%%%%%%%%%%%%%%%%%                                                                       %%%%%%%%%%%%%%%
%%%%%%%%%%%%%%%%%%%%%%%%%%%%%%%%%%%%%%%%%%%%%%%%%%%%%%%%%%%%%%%%%%%%%%%%%%%%%%%%%%%%%%%%%%%%%%%%%%%%%%%%%

\section{Generic measurement}
\label{app:rhomeasure}

The Fisher information is computed from measurement outcome probabilities and the starting point is the pre-measurement state, $\rhomeas$. The possible outcomes can be labeled by a sequence of signs. For example, $(+,-,+, \ldots )$ represents the $+$ outcome on the channel qubit, a $-$ outcome on spectator~2, a $+$ outcome on spectator~3 and so on. Then the probability that this occurs is
\begin{equation}
	\begin{split}
 P{(+,-,+, \ldots )} = 
												\frac{1}{N}
												&
												\Trace{
												\left[
													\rhomeas
													\left(
														\iop +\rinitial \cdot \sigmavec
													\right)
													\otimes
												\right.} \\
										    &
												\phantom{\Trace{}[ \rhomeas}
												\left.
													\left(
														\iop -\rinitial \cdot \sigmavec
													\right)
													\otimes\cdots
												\right].
\end{split}
\label{eq:probcalc}
\end{equation}
This requires us to determine $\rhomeas$. A simplification is that any factor of $\sigmagen{\myvector{c}}$ in $\rhomeas$ contributes zero to the probability. This results from $\Trace{[\sigmagen{a} \sigmagen{b}]} = 2 \myvector{a}^\top \myvector{b}$ and that $\rinitial$ and $\myvector{c}$ are perpendicular.

If $\rinitial = \myvector{b}$ and $\myvector{c} = \myvector{a}$, then after all channels are queried 
\begin{equation}
	\rhof = \frac{1}{N}\,
							\iop\otimes \cdots \otimes \iop
							+
							r \rhoforder{1}
\label{eq:stateafterchannels}
\end{equation}
where $\rhoforder{1}$ is given by~\cite{collins19}
\begin{eqnarray}
 \rhoforder{1} & = &
                \frac{1}{N}\,
								\left(
									\epsilon_2 \cdots \epsilon_n
								  \sigmagen{\myvector{e}} 
									\otimes
									\sigmagen{\myvector{c}}
									\otimes \cdots \otimes 
									\sigmagen{\myvector{c}}
									\right.
								\nonumber \\
							& \phantom{=} &
								\phantom{\frac{1}{N}\,}
							  +
									\mu_2 \epsilon_3 \cdots \epsilon_n
								  \sigmagen{\myvector{f}} 
									\otimes
									\rinitial \cdot \sigmavec
									\otimes
									\sigmagen{\myvector{c}}
									\otimes \cdots \otimes 
									\sigmagen{\myvector{c}}
							  +
								\cdots
								\nonumber \\
							& \phantom{=} &
								\phantom{\frac{1}{N}\,}
							  +	\epsilon_2 \epsilon_3 \cdots \mu_n
									\left.
								  \sigmagen{\myvector{f}} 
									\otimes
									\sigmagen{\myvector{c}}
									\otimes
									\sigmagen{\myvector{c}}
									\otimes \cdots \otimes 
									\rinitial \cdot \sigmavec
									\right)
								\nonumber \\				 
\label{eq:corrorderoneafterchannel}
\end{eqnarray}
where
\begin{eqnarray}
	\myvector{e} & := & M_1 \rinitial \nonumber \\
	\myvector{f} & := & M_1 \myvector{c} 
%\label{eq:}
\end{eqnarray}
and $\epsilon_j \geqslant \mu_j \geqslant \nu_j$ are ordered eigenvectors in the diagonal decomposition of
\begin{equation}
 M_j = \epsilon_j \myvector{a} \myvector{a}^\top
			+\mu_j \myvector{b} \myvector{b}^\top
			+\nu_j \myvector{d} \myvector{d}^\top. 
%\label{eq:}
\end{equation}
Note that $\sigma_j = \epsilon_j^2$ and $\tau_j = \mu_j^2.$

The action of $\uprep$ on each term of Eq.~\eqref{eq:corrorderoneafterchannel} can be evaluated by adapting Eqs.~\eqref{eq:ucaction}, giving

\begin{eqnarray}
	\uc \iop \otimes \iop \uc^\dagger									& = &	\iop \otimes \iop, 
																													\nonumber \\
	\uc \rinitial \cdot \sigmavec \otimes \iop \uc^\dagger	& = &	\rinitial \cdot \sigmavec \otimes \sigmagen{c}
																													\nonumber \\
	\uc \rinitial \cdot \sigmavec \otimes \sigmagen{c} \uc^\dagger	& = &	\rinitial \cdot \sigmavec \otimes \iop
																													\nonumber \\
	\uc \sigmagen{b} \otimes \iop \uc^\dagger	& = &	\sigmagen{b} \otimes \sigmagen{c}  
																													\nonumber \\
																										& 	& + \left( \myvector{b} \cdot \myvector{c} \right)
																													\left(
																														 \sigmagen{c} \otimes \iop - \sigmagen{c} \otimes \sigmagen{c}
																													\right), 
																													\nonumber \\
	\uc \sigmagen{b} \otimes \sigmagen{c} \uc^\dagger	& = &	\sigmagen{b} \otimes \iop  
																													\nonumber \\
																										& 	& -\left( \myvector{b} \cdot \myvector{c} \right)
																													\left(
																														 \iop \otimes \sigmagen{c} - \sigmagen{c} \otimes \sigmagen{c}
																													\right), 
																													\;
																													\textrm{and}
																													\nonumber \\
	\uc \sigmagen{b} \otimes \rinitial \cdot \sigmavec \uc^\dagger	& = &	\left( \myvector{b} \times \myvector{c} \right)
																													\cdot \hat{\myvector{\sigma}} \otimes 
																													\left( \myvector{b} \times \rinitial \right)
																													\cdot
																													\hat{\myvector{\sigma}}
																													\nonumber \\
																										& 	& \left( \myvector{b} \cdot \myvector{c} \right)
																													\iop \otimes \rinitial \cdot \sigmavec.
	\label{eq:ucactionro}
\end{eqnarray}

First, this maps the identity term in Eq.~\eqref{eq:stateafterchannels} to itself. Then Eq.~\eqref{eq:probcalc} shows that this contributes $1/N$ to the probability regardless of the measurement outcome. 

Second consider the action of $\uprep$ on the term 
$\sigmagen{\myvector{e}} 
\otimes
\sigmagen{\myvector{c}}
\otimes \cdots \otimes 
\sigmagen{\myvector{c}}$
in Eq.~\eqref{eq:corrorderoneafterchannel}. The only factors of $\uc$ in $\uprep$ that alter this are those involving the leftmost qubit. According to Eq.~\eqref{eq:ucactionro} these produce
$\sigmagen{\myvector{e}} 
\otimes
\iop
\otimes \cdots \otimes 
\iop
$
plus terms that contain a factor of $\sigmagen{c}$ and which do not contribute to the probability. Then Eq.~\eqref{eq:probcalc} shows that this contributes $\pm r (\rinitial^\top M_1 \rinitial)\, \epsilon_2 \cdots \epsilon_n/N$ to the probability $P(\pm,\ldots).$  

Third, consider the action of $\uprep$ on the term 
$\sigmagen{\myvector{f}} 
\otimes
\rinitial \cdot \sigmavec 
\otimes
\sigmagen{\myvector{c}}
\otimes \cdots \otimes 
\sigmagen{\myvector{c}}$
in Eq.~\eqref{eq:corrorderoneafterchannel}. Equation~\eqref{eq:ucactionro} shows that the factors of $\uc$ that act on all but the leftmost qubit map this to
$\sigmagen{\myvector{f}}
\otimes
\rinitial \cdot \sigmavec 
\otimes
\iop
\otimes \cdots \otimes 
\iop.$
Then $\uc$ acting on the leftmost two qubits maps this to
$
\left(\myvector{c}^\top M_1 \myvector{c}\right)\iop
\otimes
\rinitial \cdot \sigmavec 
\otimes
\iop
\otimes \cdots \otimes 
\iop$
along with terms that have a factor of $[(M_1 \myvector{c}) \times \rinitial]\cdot \sigmavec$ for the second to left qubit (this will trace to zero). Then Eq.~\eqref{eq:probcalc} shows that this contributes $\pm r (\myvector{c}^\top M_1 \myvector{c})\, \mu_2 \epsilon_3 \cdots \epsilon_n/N$ to the probability $P(.,\pm,\ldots).$

Collecting these gives
\begin{widetext}
\begin{eqnarray}
 P(+,+,\ldots,+) & = & \frac{1}{N}\,
											\left\{
												1 + 
												r
												\left[
													(\rinitial^\top M_1 \rinitial)\, \epsilon_2 \cdots \epsilon_n
													+
													(\myvector{c}^\top M_1 \myvector{c})\, 
														\left(
														 \mu_2 \epsilon_3\cdots \epsilon_n
														+\epsilon_2 \mu_3 \epsilon_4 \cdots \epsilon_n
														+ \cdots
														  +
														 \epsilon_2 \cdots \epsilon_{n-1} \mu_n
														\right)
												\right]
											\right\}
											\nonumber \\
 P(+,+,\ldots,+,-) & = & \frac{1}{N}\,
											\left\{
												1 + 
												r
												\left[
													(\rinitial^\top M_1 \rinitial)\, \epsilon_2 \cdots \epsilon_n
													+
													(\myvector{c}^\top M_1 \myvector{c})\, 
														\left(
														 \mu_2 \epsilon_3\cdots \epsilon_n
														+\epsilon_2 \mu_3 \epsilon_4 \cdots \epsilon_n
														+ \cdots
														-\epsilon_2 \cdots \epsilon_{n-1} \mu_n
														\right)
												\right]
											\right\}
											\nonumber \\
								& \vdots & 
											\nonumber \\
 P(-,-,\ldots,-) & = & \frac{1}{N}\,
											\left\{
												1 + 
												r
												\left[
													-(\rinitial^\top M_1 \rinitial)\, \epsilon_2 \cdots \epsilon_n
													+
													(\myvector{c}^\top M_1 \myvector{c})\, 
														\left(
														 -\mu_2 \epsilon_3\cdots \epsilon_n
														-\epsilon_2 \mu_3 \epsilon_4 \cdots \epsilon_n
														- \cdots
														-\epsilon_2 \cdots \epsilon_{n-1} \mu_n
														\right)
												\right]
											\right\}.
											\nonumber \\
\label{eq:genericmeasprobs}
\end{eqnarray}
\end{widetext}

The contributions of these to the classical Fisher information
\begin{equation}
 F = \sum_\textrm{outcomes $m$} \frac{1}{P_m} \left( \frac{\partial P_m}{\partial \lambda} \right)^2
%\label{eq:}
\end{equation}
can be assessed by pairing the terms in the sum. For example, consider $P(+,+,\ldots,+)$ and $P(-,-,\ldots,-)$. Equations~\eqref{eq:genericmeasprobs} show
\begin{equation}
	\left( \frac{\partial P(+,+,\ldots,+)}{\partial \lambda} \right)^2
	=
	\left( \frac{\partial P(-,-,\ldots,)}{\partial \lambda} \right)^2.
%\label{eq:}
\end{equation}
Thus these two terms contribute
\begin{equation}
	\left( \frac{1}{P(+,+,\ldots,+)} + \frac{1}{P(-,-,\ldots,-)}\right)
	\left( \frac{\partial P(+,+,\ldots,+)}{\partial \lambda} \right)^2
%\label{eq:}
\end{equation}
to the Fisher information. To lowest order in the purity, $\frac{1}{P(+,+,\ldots,+)} + \frac{1}{P(-,-,\ldots,-)} = 2N$ while the derivative is proportional to $r$. Thus, to lowest order in the purity, these terms contribute
\begin{widetext}
\begin{equation}
	\frac{2 r^2}{N}\,
	\left[
		(\rinitial^\top \dot{M}_1 \rinitial)\, \epsilon_2 \cdots \epsilon_n
													+
													(\myvector{c}^\top \dot{M}_1 \myvector{c})\, 
														\left(
														 \mu_2 \epsilon_3\cdots \epsilon_n
														+\epsilon_2 \mu_3 \epsilon_4 \cdots \epsilon_n
														+ \cdots
														  +
														 \epsilon_2 \cdots \epsilon_{n-1} \mu_n
														\right)
	\right]^2
%\label{eq:}
\end{equation}
to the Fisher information. Extending this to the remaining terms gives
\begin{eqnarray}
 F & = & 	\frac{2 r^2}{N}\,
					\left\{ 
						\left[
							(\rinitial^\top \dot{M}_1 \rinitial)\, \epsilon_2 \cdots \epsilon_n
																		+
																		(\myvector{c}^\top \dot{M}_1 \myvector{c})\, 
																			\left(
																			 \mu_2 \epsilon_3\cdots \epsilon_n
																			+\epsilon_2 \mu_3 \epsilon_4 \cdots \epsilon_n
																			+ \cdots
																				+
																			 \epsilon_2 \cdots \epsilon_{n-1} \mu_n
																			\right)
						\right]^2
					\right.
					\nonumber \\
		& & + 
						\left[
							-(\rinitial^\top \dot{M}_1 \rinitial)\, \epsilon_2 \cdots \epsilon_n
																		+
																		(\myvector{c}^\top \dot{M}_1 \myvector{c})\, 
																			\left(
																			 \mu_2 \epsilon_3\cdots \epsilon_n
																			+\epsilon_2 \mu_3 \epsilon_4 \cdots \epsilon_n
																			+ \cdots
																				+
																			 \epsilon_2 \cdots \epsilon_{n-1} \mu_n
																			\right)
						\right]^2
						+ \cdots 
					\nonumber \\
		& & + 	\left.
						\left[
							-(\rinitial^\top \dot{M}_1 \rinitial)\, \epsilon_2 \cdots \epsilon_n
																		+
																		(\myvector{c}^\top \dot{M}_1 \myvector{c})\, 
																			\left(
																			 -\mu_2 \epsilon_3\cdots \epsilon_n
																			-\epsilon_2 \mu_3 \epsilon_4 \cdots \epsilon_n
																			- \cdots
																			-
																			 \epsilon_2 \cdots \epsilon_{n-1} \mu_n
																			\right)
						\right]^2
						\right\}.
%\label{eq:}
\end{eqnarray}
Every combination of $\pm$ signs occurs exactly once in this expression. This implies that all the cross terms after squaring will cancel. Thus only terms such as $[(\rinitial^\top \dot{M}_1 \rinitial)\, \epsilon_2 \cdots \epsilon_n]^2$ contribute to F. There are exactly $N/2$ of each. Thus
\begin{equation}
 F = r^2 	\left\{
						(\rinitial^\top \dot{M}_1 \rinitial)^2\, 
						(\epsilon_2 \cdots \epsilon_n)^2
						+
						(\myvector{c}^\top \dot{M}_1 \myvector{c})^2\, 
							\left[
							 \left(\mu_2 \epsilon_3\cdots \epsilon_n \right)^2
							+\left(\epsilon_2 \mu_3 \epsilon_4 \cdots \epsilon_n\right)^2
							+ \cdots
							+
							\left(\epsilon_2 \cdots \epsilon_{n-1} \mu_n \right)^2
							\right]
						\right\}.
%\label{eq:}
\end{equation}
\end{widetext}
Finally $\sigma_j = \epsilon_j^2$ and $\tau_j = \mu_j^2$ give the result of Eq.~\eqref{eq:Fgenericmeas}.

%%%%%%%%%%%%%%%%%%%%%%%%%%%%%%%%%%%%%%%%%%%%%%%%%%%%%%%%%%%%%%%%%%%%%%%%%%%%%%%%%%%%%%%%%%%%%%%%%%%%%%%%%
%%%%%%%%%%%%%%%%%%%                                                                       %%%%%%%%%%%%%%%
%%%%%%%%%%%%%%%%%%%         Begin section                                                 %%%%%%%%%%%%%%%
%%%%%%%%%%%%%%%%%%%                                                                       %%%%%%%%%%%%%%%
%%%%%%%%%%%%%%%%%%%%%%%%%%%%%%%%%%%%%%%%%%%%%%%%%%%%%%%%%%%%%%%%%%%%%%%%%%%%%%%%%%%%%%%%%%%%%%%%%%%%%%%%%

\section{Tailored measurement scheme}
\label{app:rhomeasurestailored}

The reasoning of Appendix~\ref{app:rhomeasure} applies here. The state after the channels is given by Eqs.~\eqref{eq:stateafterchannels} and~\eqref{eq:corrorderoneafterchannel}. Here the factors of $\uc$ only act on the rightmost $n-1$ qubits. 

First, the identity term in Eq.~\eqref{eq:stateafterchannels} maps to itself. This again contributes $1/N$ to the probability. 

Second, the term 
$\sigmagen{\myvector{e}} 
\otimes
\sigmagen{\myvector{c}}
\otimes \cdots \otimes 
\sigmagen{\myvector{c}}$
in Eq.~\eqref{eq:corrorderoneafterchannel} maps to itself. The factors of $\sigmagen{c}$ results in zero contribution to the probability. 

Third, the term 
$\sigmagen{\myvector{f}} 
\otimes
\rinitial \cdot \sigmavec 
\otimes
\sigmagen{\myvector{c}}
\otimes \cdots \otimes 
\sigmagen{\myvector{c}}$
in Eq.~\eqref{eq:corrorderoneafterchannel} maps to
$\sigmagen{\myvector{f}}
\otimes
\rinitial \cdot \sigmavec 
\otimes
\iop
\otimes \cdots \otimes 
\iop.$
The contributes $r^2 (\myvector{m}^\top M_1 \myvector{c}) \mu_2 \epsilon_3\ldots \epsilon_n/N$ to the probability. 

Continuing this analysis gives
\begin{widetext}
\begin{eqnarray}
 P(+,+,\ldots,+) & = & \frac{1}{N}\,
											\left\{
												1 + 
												r
													(\myvector{m}^\top M_1 \myvector{c})\, 
														\left(
														 \mu_2 \epsilon_3\cdots \epsilon_n
														+\epsilon_2 \mu_3 \epsilon_4 \cdots \epsilon_n
														+ \cdots
														  +
														 \epsilon_2 \cdots \epsilon_{n-1} \mu_n
														\right)
											\right\}
											\nonumber \\
 P(+,+,\ldots,+,-) & = & \frac{1}{N}\,
											\left\{
												1 + 
												r
													(\myvector{m}^\top M_1 \myvector{c})\, 
														\left(
														 \mu_2 \epsilon_3\cdots \epsilon_n
														+\epsilon_2 \mu_3 \epsilon_4 \cdots \epsilon_n
														+ \cdots
														 -
														 \epsilon_2 \cdots \epsilon_{n-1} \mu_n
														\right)
											\right\}
											\nonumber \\
								& \vdots & 
											\nonumber \\
 P(-,-,\ldots,-) & = & \frac{1}{N}\,
											\left\{
												1 + 
												r
													(\myvector{m}^\top M_1 \myvector{c})\, 
														\left(
														- \mu_2 \epsilon_3\cdots \epsilon_n
														- \epsilon_2 \mu_3 \epsilon_4 \cdots \epsilon_n
														- \cdots
														- \epsilon_2 \cdots \epsilon_{n-1} \mu_n
														\right)
											\right\}.
											\nonumber \\
\label{eq:genericmeasprobstwo}
\end{eqnarray}
\end{widetext}

The Fisher information is determined in the same fashion as for the generic measurement case and the same arguments offered in Appendix~\ref{app:rhomeasure} give the result of Eq.~\eqref{eq:Ftailoredmeas}.

%\bibliography{../biblio/refs}


\end{document}